\documentclass[a4paper]{myjpconf}
\usepackage{graphicx,amssymb}

\begin{document}
\title{On the equivalence of the Pais-Uhlenbeck oscillator model and two non-Hermitian Harmonic Oscillators}

\author{Frieder Kleefeld}

\address{Collab. of CeFEMA, IST, Lisbon, Portugal\\
Address for correspondence: Frankenstr.\ 3, D-91452 Wilhermsdorf, Germany}

\ead{frieder.kleefeld@web.de}

\begin{abstract}
A system of two independent Bosonic Harmonic Oscillators is converted into  the respective  fourth-order derivative Pais-Uhlenbeck  oscillator model. The conversion procedure displays transparently how the quantization of the  fourth-order derivative Pais-Uhlenbeck  oscillator has to be performed in order not to suffer from the divergence problems of the vacuum state and path integrals as conjectured most recently by P.\ D.\ Mannheim in his article ``Determining the normalization of the quantum field theory vacuum, with implications for quantum gravity" [arXiv:2301.13029 [hep-th]]. In order to make the case we present the construction of the path integral, generating functionals and vacuum persistence amplitudes for  PT-symmetry completed systems in Quantum Mechanics and Quantum Field Theory  and discuss some implications to Quantum Field Theory and Particle Physics.
\end{abstract}

\section{Introduction}
Due to its implications for the construction of a renormalizable Quantum Theory of Gravity \cite{Mannheim:2019aap}\cite{Mannheim:2023ivd} and the quantization of higher derivative theories \cite{deUrries:1998obu}\cite{Smilga:2017arl}\cite{Mandal:2022xuw} the fourth-order derivative Pais-Uhlenbeck \cite{Pais:1950za} oscillator model \cite{Mannheim:2023ivd}\cite{Mandal:2022xuw}\cite{Bender:2007wu}\cite{Mannheim:2004qz} is receiving more and more interest. Most recently it has been argued by P.\ D.\ Mannheim \cite{Mannheim:2023ivd} that the formulation of Minkowskian or Euclidean path integrals for such models should be done with due care due to a problem with the normalizability of the vacuum states being related to to the non-Hermitian nature of the formalism. 

In this manuscript we would like to take the reverse point of view in demonstrating that a --- seemingly Hermitian --- system of two independent Bosonic Harmonic Oscillators can be converted --- under certain conditions --- straight forwardly into the respective  fourth-order derivative Pais-Uhlenbeck  oscillator model under consideration e.\ g.\ by P.\ D.\ Mannheim. As the system of two independent Harmonic Oscillators does --- after PT-symmetric completion --- not suffer from the problem of a diverging norm of the vacuum state and diverging path integrals when being quantized canonically, it gets clear that also the respective  fourth-order derivative Pais-Uhlenbeck  oscillator model must be free of these problems when quantized appropriately.

\section{Deriving the Pais-Uhlenbeck oscillator model from two Harmonic Oscillators}
As a starting point we consider the degrees of freedom $Q_1(t)$ and $Q_2(t)$ of two independent Harmonic oscillators with angular frequencies $\Omega_1$ and $\Omega_2$ respecting the following equations of motion being differential equations of second oder in time:
\begin{equation} \ddot{Q}_1(t)  \,  + \,  \Omega^2_1 \,  Q_1(t)  = 0 \; , \qquad \ddot{Q}_2(t)  \,  + \,  \Omega^2_2 \,  Q_2(t)  = 0 \; .\end{equation} 
It is well known that the solutions of these linear second order differential equations can be decomposed into a sum of two independent solutions, i.\ e.,
\begin{eqnarray} Q_{1} (t) & = & Q^{(+)}_1 (t) \; + \; Q^{(-)}_1 (t) \; = \;   Q^{(+)}_1 (0) \; e^{-\,i\,\Omega_1 t}\; + \; Q^{(-)}_1 (0) \; e^{+\,i\,\Omega_1 t} \; , \nonumber \\[2mm]
 Q_{2} (t) & = & Q^{(+)}_2 (t) \; + \; Q^{(-)}_2 (t) \; = \;   Q^{(+)}_2 (0) \; e^{-\,i\,\Omega_2 t}\; + \; Q^{(-)}_2 (0) \; e^{+\,i\,\Omega_2 t} \; . \label{qsolut1}
 \end{eqnarray}
The Lagrange function describing the system of two independent Bosonic Harmonic Oscillators of mass $M_1$ and $M_2$, respectively, is well known, i.\ e.,
\begin{equation} L_H[Q_1,Q_2,\dot{Q}_1,\dot{Q}_2] \; = \; \frac{M_1}{2} \left( \dot{Q}_{1} (t)^2 \, - \, \Omega^2_1  \, Q_1(t)^2\right) \; + \; \frac{M_2}{2} \left( \dot{Q}_{2} (t)^2 \, - \, \Omega^2_2  \, Q_2(t)^2\right) \; . \label{lagtwoho1}
\end{equation}
The differential equation fourth-order in time describing the  Pais-Uhlenbeck  oscillator model discussed e.\ g.\ in \cite{Mannheim:2023ivd}\cite{Smilga:2017arl}\cite{Bender:2007wu} is given by:
\begin{equation} \left( \frac{d^2}{dt^2} \, + \, \Omega^2_1 \right) \left( \frac{d^2}{dt^2} \, + \, \Omega^2_2 \right) \, Q(t) \; = \; 0 \; . \end{equation}
The degree of freedom $Q(t)$ solving this linear differential equation can be obviously decomposed into a sum of degrees of freedom describing the two independent Bosonic Harmonic Oscillators, i.\ e.,
\begin{eqnarray} Q(t) & = & Q_1(t) + Q_2(t) \; = \; Q^{(+)}_1 (t) \; + \; Q^{(-)}_1 (t) \; + \; Q^{(+)}_2 (t) \; + \; Q^{(-)}_2 (t) \nonumber \\[2mm]
 & = & Q^{(+)}_1 (0) \; e^{-\,i\,\Omega_1 t}\; + \; Q^{(-)}_1 (0) \; e^{+\,i\,\Omega_1 t} \; + \; Q^{(+)}_2 (0) \; e^{-\,i\,\Omega_2 t}\; + \; Q^{(-)}_2 (0) \; e^{+\,i\,\Omega_2 t}  \; .  \end{eqnarray}
It is straight forward to show that there hold the following identities:
\begin{eqnarray} \left( \begin{array}{c} Q(t) \\[2mm] \ddot{Q} (t)\end{array}\right) & = & \left( \begin{array}{cc} 1 & 1 \\[2mm] -\,\Omega_1^2 & -\,\Omega^2_2\end{array}\right) \left( \begin{array}{c} Q_1(t) \\[2mm] Q_2 (t)\end{array}\right) \; , \\[2mm]
\left( \begin{array}{c} \dot{Q}(t) \\[2mm] \stackrel{...}{Q} \!\!(t)\end{array}\right) & = & \left( \begin{array}{cc} 1 & 1 \\[2mm] -\,\Omega_1^2 & -\,\Omega^2_2\end{array}\right) \left( \begin{array}{c} \dot{Q}_1(t) \\[2mm] \dot{Q}_2 (t)\end{array}\right) \; ,
\end{eqnarray}
which can be easily inverted, i.\ e.,
\begin{eqnarray}  \left( \begin{array}{c} Q_1(t) \\[2mm] Q_2 (t)\end{array}\right)  & = & \frac{1}{\Omega_1^2 \, - \, \Omega_2^2} \left( \begin{array}{rr} -\, \Omega_2^2 & -\, 1 \\[2mm] \Omega_1^2 & 1\end{array}\right)\left( \begin{array}{c} Q(t) \\[2mm] \ddot{Q} (t)\end{array}\right) \; , \\[2mm]
\left( \begin{array}{c} \dot{Q}_1(t) \\[2mm] \dot{Q}_2 (t)\end{array}\right) & = &  \frac{1}{\Omega_1^2 \, - \, \Omega_2^2} \left( \begin{array}{rr} -\, \Omega_2^2 & -\, 1 \\[2mm] \Omega_1^2 & 1\end{array}\right) \left( \begin{array}{c} \dot{Q}(t) \\[2mm] \stackrel{...}{Q}\!\! (t)\end{array}\right) \; ,
\end{eqnarray}
or, equivalently, 
\begin{eqnarray} Q_1(t) & = & \frac{\Omega_2^2 \, Q(t) + \ddot{Q}(t)}{\Omega_2^2\, - \, \Omega_1^2} \; , \qquad Q_2(t) \: = \; \frac{\Omega_1^2 \, Q(t) + \ddot{Q}(t)}{\Omega_1^2\, - \, \Omega_2^2} \; , \\[2mm]
\dot{Q}_1(t) & = & \frac{\Omega_2^2 \, \dot{Q}(t) \,+ \,\stackrel{...}{Q}\!\!(t)}{\Omega_2^2\, - \, \Omega_1^2} \; , \qquad \dot{Q}_2(t) \: = \; \frac{\Omega_1^2 \, \dot{Q}(t)\, +\, \stackrel{...}{Q}\!\!(t)}{\Omega_1^2\, - \, \Omega_2^2} \; . \end{eqnarray}
By inserting these expressions into the Lagrange function for the system of two independent Bosonic Harmonic Oscillators Eq.\ (\ref{lagtwoho1}) we obtain some suitable Lagrange function for the respective fourth-order derivative Pais-Uhlenbeck  oscillator model, i.\ e., 
\begin{eqnarray} L_H[Q,\dot{Q},\ddot{Q},\stackrel{...}{Q}] & = & \frac{1}{2 \,(\Omega_1^2\, - \, \Omega_2^2)^2} \;  \Bigg\{  M_1  \, \Big[ \left( \Omega_2^2 \, \dot{Q}(t) \,+ \stackrel{...}{Q}\!\!(t)\right)^2 \, - \, \Omega^2_1  \, \left(\Omega_2^2 \, Q(t) + \ddot{Q}(t)\right)^2\Big] \nonumber \\[1mm]
 &  & \qquad\qquad\quad \;\;+ \; M_2\, \Big[ \left( \Omega_1^2 \, \dot{Q}(t) \,+ \stackrel{...}{Q}\!\!(t)\right)^2 \, - \, \Omega^2_2  \, \left(\Omega_1^2 \, Q(t) + \ddot{Q}(t)\right)^2\Big] \Bigg\} \; . \quad\;\; \label{lagpaisuhl1}   
\end{eqnarray}
or, equivalently,
\begin{eqnarray} \lefteqn{L_H[Q,\dot{Q},\ddot{Q},\stackrel{...}{Q}] \; = \; \frac{1}{2 \,(\Omega_1^2\, -\, \Omega_2^2)^2} \; \cdot} \nonumber \\
 & \cdot &   \Bigg\{ \, M_1 \Big[\, \Omega_2^4  \left( \dot{Q}(t)^2 -  \Omega_1^2 \, Q(t)^2\right) \! +  \stackrel{...}{Q}\!\!(t)^2  \!-  \Omega^2_1  \,  \ddot{Q}(t)^2\!+  \Omega_2^2  \left( \{ \dot{Q}(t) , \stackrel{...}{Q}\!\!(t)\} \! - \Omega_1^2 \, \{Q(t),  \ddot{Q}(t)\}\right) \Big] \nonumber \\[1mm]
 &  & + \; M_2 \Big[ \, \Omega_1^4  \left( \dot{Q}(t)^2 - \Omega_2^2 \, Q(t)^2\right) \!+  \stackrel{...}{Q}\!\!(t)^2 \! -  \Omega^2_2  \,  \ddot{Q}(t)^2\!+ \Omega_1^2  \left( \{\dot{Q}(t) , \stackrel{...}{Q}\!\!(t)\} \! - \Omega_2^2 \, \{ Q(t) , \ddot{Q}(t)\}\right)  \Big] \!\Bigg\} \nonumber \\
 & = & \frac{1}{2 \,(\Omega_1^2\, -\, \Omega_2^2)^2} \; \cdot \nonumber \\
 & \cdot &   \Bigg\{ \, M_1 \, \Big[  -  \Omega_1^2\, \Omega_2^4 \, Q(t)^2  \, +  \stackrel{...}{Q}\!\!(t)^2  -  \left( \Omega^2_1+ 2\, \Omega_2^2 \right)  \ddot{Q}(t)^2  + \Omega_2^2 \left( 2\,  \Omega_1^2 +  \Omega_2^2\right) \dot{Q}(t)^2  \nonumber \\
 &  &  \quad\;\;\;\; \;\;\;\;\,+\,   \Omega_2^2 \;\, \frac{d}{dt}\! \left( \{ \dot{Q}(t) , \ddot{Q}(t)\}  - \,\Omega_1^2 \, \{Q(t),  \dot{Q}(t)\}\right) \Big] \nonumber \\[1mm]
 &  & + \; M_2 \,\Big[ - \Omega_1^4  \, \Omega_2^2 \, Q(t)^2 \,+  \stackrel{...}{Q}\!\!(t)^2 - \left(  2\,  \Omega_1^2 +  \Omega^2_2\right) \ddot{Q}(t)^2  +\Omega_1^2 \left(  \Omega_1^2 + 2\,   \Omega_2^2\right)  \dot{Q}(t)^2  \nonumber \\
 &  &  \quad\;\;\;\;\;\;\;\;\,+ \,  \Omega_1^2 \;\, \frac{d}{dt}\! \left( \{\dot{Q}(t) , \ddot{Q}(t)\}  - \,\Omega_2^2 \, \{ Q(t) , \dot{Q}(t)\}\right)  \Big] \Bigg\}\, .   \label{lagpaisuhl2} 
\end{eqnarray}
During the last step we have invoked  partial integrations.
In using the identities $M_1=\frac{1}{2}(M_1+M_2) + \frac{1}{2}(M_1-M_2)$ and $M_2=\frac{1}{2}(M_1+M_2) - \frac{1}{2}(M_1-M_2)$ the result can be transformed to the following convenient form:

\begin{eqnarray} \lefteqn{L_H[Q,\dot{Q},\ddot{Q},\stackrel{...}{Q}] \; =} \nonumber \\[3mm]
 & = & \frac{1}{2 \,(\Omega_1^2\, -\, \Omega_2^2)^2} \; \cdot \nonumber \\[3mm]
 & \cdot &   \Bigg\{ \, \frac{M_1+M_2}{2} \, \Big[ \left(\Omega_1^4+\Omega_2^4 \right)\, \dot{Q}(t)^2 +  \left(\Omega^2_1 +  \Omega^2_2\right) \left(-\,  \ddot{Q}(t)^2 -  \Omega_1^2\,\Omega_2^2 \, Q(t)^2 +    \{ \dot{Q}(t) , \stackrel{...}{Q}\!\!(t)\} \right)   \nonumber \\[1mm]
 &  & \qquad\qquad\;\;\,\,+ \,2 \left(  \stackrel{...}{Q}\!\!(t)^2  -  \Omega_1^2\,\Omega_2^2 \, \{ Q(t) , \ddot{Q}(t)\} \right)  \! \Big]  \nonumber \\
 &  &  +  \, \frac{M_1-M_2}{2}   \left(\Omega_1^2 - \Omega_2^2\right)  \Big[-\ddot{Q}(t)^2 -\left(\Omega_1^2 + \Omega_2^2\right)  \dot{Q}(t)^2 +  \Omega_1^2 \, \Omega_2^2 \, Q(t)^2 - \{ \dot{Q}(t) , \stackrel{...}{Q}\!\!(t)\}  \Big]  \Bigg\} \nonumber \\
 & = & \frac{1}{2 \,(\Omega_1^2\, -\, \Omega_2^2)^2} \; \cdot \nonumber \\[3mm]
 & \cdot &   \Bigg\{ \, \frac{M_1+M_2}{2} \, \Big[  - \left( \Omega^2_1+  \Omega_2^2 \right) \left(3\, \ddot{Q}(t)^2+ \Omega_1^2\, \Omega_2^2 \, Q(t)^2    \right)  +\left(  \Omega_1^4 +\Omega_2^4 + 4\,  \Omega_1^2 \, \Omega_2^2\right)  \dot{Q}(t)^2  \nonumber \\[1mm]
 &  &  \quad\quad\quad\quad \;\;\;\; \;\;\;\,   +  \, 2 \stackrel{...}{Q}\!\!(t)^2 + \,  \frac{d}{dt}\! \left( \left(\Omega_1^2 +\Omega_2^2 \right) \{\dot{Q}(t) , \ddot{Q}(t)\}  - \,2 \,\Omega_1^2 \,\Omega_2^2 \, \{ Q(t) , \dot{Q}(t)\}\right)  \Big] \nonumber \\[2mm]
 &  &  +  \, \frac{M_1-M_2}{2}   \left(\Omega_1^2 - \Omega_2^2\right)  \Big[\ddot{Q}(t)^2 -\left(\Omega_1^2 + \Omega_2^2\right)  \dot{Q}(t)^2 +  \Omega_1^2 \, \Omega_2^2 \, Q(t)^2 - \frac{d}{dt}  \{\dot{Q}(t) , \ddot{Q}(t)\}  \Big]  \Bigg\}   \, . \nonumber \\  \label{lagpaisuhl3} 
\end{eqnarray}
The result can be compared to the following expression for the fourth-order derivative Pais-Uhlenbeck  oscillator model  considered e.\ g.\ by C.\ M. Bender and P.\ D.\ Mannheim and rewritten here using our notation:
\begin{equation} L_{H} \; = \; \frac{\gamma}{2} \left[ \, \ddot{Q}(t)^2 -\left(\Omega_1^2 + \Omega_2^2\right)  \dot{Q}(t)^2 +  \Omega_1^2 \, \Omega_2^2 \, Q(t)^2 \, \right] \; . \label{lagbenman1}\end{equation} 
A comparison of Eqs.\ (\ref{lagpaisuhl3}) and (\ref{lagbenman1}) reveals that C.\ M. Bender and P.\ D.\ Mannheim \cite{Bender:2007wu} consider the particular case $M_1 = -M_2$   implying
\begin{equation} \gamma = \frac{M_1}{\Omega_1^2\, -\, \Omega_2^2} \; = \;  \frac{-\, M_2}{\Omega_1^2\, -\, \Omega_2^2} \; . \end{equation}
Although $M_1$ and $M_2$ can take here arbitrary non-vanishing even complex values we will consider here without loss of generality the mass parameter $M_1$ to be positive yielding $M_2$ to be negative. 
\section{The non-Hermitian nature of seemingly Hermitian  Harmonic Oscillators}
Since the standard quantization of the system of two independent Bosonic Harmonic Oscillators will lead to the following well known energy eigen-values
\begin{equation} E_{m_1,\,m_2} \; = \; \hbar \, \Omega_1 \left( m_1 + \frac{1}{2}\right) +  \hbar \,\Omega_2 \left( m_2 + \frac{1}{2}\right) \label{spec1} \end{equation}
with $m_1$ and $m_2$ being arbitrary non-negative integers  it might appear as a surprise that the spectrum  of two independent Bosonic Harmonic Oscillators will not be affected by the fact that the mass parameters $M_1$ and $M_2$ of the two Harmonic Oscillators have opposite sign. In order to explain this puzzling point we first note by inspecting Eq.\ (\ref{spec1}) that the spectrum merely depends on the frequencies $\Omega_1$ and $\Omega_2$ and it does not at all depend on the choice of the mass parameters $M_1$ and $M_2$. The stationary Schr\"odinger equation possessing the spectrum of Eq.~(\ref{spec1}) for square-integrable eigen-functions $\phi(\rho_1,\rho_2)$ of two dimensionless  real-valued arguments $\rho_1$ and $\rho_2$  is given by
\begin{equation} \left[ \, \frac{\hbar \, \Omega_1}{2} \left( -\, \frac{d^2}{d\rho^2_1} + \rho^2_1 \right) +  \frac{\hbar \,\Omega_2}{2} \left(  -\, \frac{d^2}{d\rho^2_2} + \rho^2_2\right) \, \right] \, \phi_{m_1,m_2} (\rho_1,\rho_2)\; = \;  E_{m_1,\,m_2} \; \phi_{m_1,m_2} (\rho_1,\rho_2) \; . \label{schroed1} \end{equation}
The respective orthonormality condition of the eigen-functions reads:
\begin{equation} \int\limits^{+\infty}_{-\infty} \! d\rho_1 \int\limits^{+\infty}_{-\infty} \! d\rho_2 \; \; \phi^\ast_{m_1,\,m_2} (\rho_1,\rho_2) \; \phi_{m^\prime_1,\,m^\prime_2} (\rho_1,\rho_2) \; = \; \delta_{m_1,\,m_1^\prime}  \; \delta_{m_2,\,m_2^\prime} \; . \end{equation}
It is well known that the Fourier-transform of a square-integrable function is also square-integrable. The Fourier-transform $\chi_{m^\prime_1,\,m^\prime_2}(\kappa_1,\kappa_2)$  of the eigen-functions $\phi_{m_1,\,m_2} (\rho_1,\rho_2)$ being also functions of dimensionless real-valued arguments $\kappa_1$ and $\kappa_2$ are obtained by
\begin{eqnarray} \chi_{m_1,\,m_2}(\kappa_1,\kappa_2) & = & \int\limits^{+\infty}_{-\infty} \! d\rho_1 \int\limits^{+\infty}_{-\infty} \! d\rho_2 \; \; \phi_{m_1,\,m_2} (\rho_1,\rho_2) \; e^{-\,i\,(\kappa_1\rho_1+\kappa_2\rho_2)} \; ,
\end{eqnarray}
or, inversely,
\begin{eqnarray} \phi_{m_1,\,m_2} (\rho_1,\rho_2) & = & \int\limits^{+\infty}_{-\infty} \! \frac{d\kappa_1}{2\pi} \int\limits^{+\infty}_{-\infty} \! \frac{d\kappa_2}{2\pi} \; \; \chi_{m_1,\,m_2}(\kappa_1,\kappa_2) \; e^{+\,i\,(\kappa_1\rho_1+\kappa_2\rho_2)}  \; .
\end{eqnarray}
The Fourier-transformed eigen-functions $\chi_{m^\prime_1,\,m^\prime_2} (\kappa_1,\kappa_2)$ are also orthonormal, i.\ e.,
\begin{equation} \int\limits^{+\infty}_{-\infty} \! \frac{d\kappa_1}{2\pi} \int\limits^{+\infty}_{-\infty} \! \frac{d\kappa_2}{2\pi} \; \; \chi^\ast_{m_1,\,m_2} (\kappa_1,\kappa_2) \; \chi_{m^\prime_1,\,m^\prime_2} (\kappa_1,\kappa_2) \; = \; \delta_{m_1,\,m_1^\prime}  \; \delta_{m_2,\,m_2^\prime} \; , \end{equation}
and are the eigen-solutions of the following Fourier-transformed Schr\"odinger equation \cite{Fluegge:1994}:
\begin{equation} \left[ \, \frac{\hbar \, \Omega_1}{2} \left(  \kappa^2_1 -\, \frac{d^2}{d\kappa^2_1} \right) +  \frac{\hbar \,\Omega_2}{2} \left(  \kappa^2_2-\, \frac{d^2}{d\kappa^2_2}  \right) \, \right] \, \chi_{m_1,m_2} (\kappa_1,\kappa_2)\; = \;  E_{m_1,\,m_2} \; \chi_{m_1,m_2} (\kappa_1,\kappa_2) \; . \label{fschroed1} \end{equation}
In terms of the Harmonic Oscillator degrees of freedom $Q_1(t)$ and $Q_2(t)$ (or respectively $\zeta_1$ and $\zeta_2$) and their canonical conjugate momenta $P_1(t)$ and $P_2(t)$ (or respectively $p_1$ and $p_2$) the Schr\"odinger Eqs.\ (\ref{schroed1}) and (\ref{fschroed1}) will have to be denoted in the following way:
\begin{equation} \left[ \, - \frac{\hbar^2}{2 M_1} \frac{d^2}{d\zeta^2_1} + \frac{M_1\Omega^2_1}{2} \;  \zeta^2_1   - \frac{\hbar^2}{2 M_2} \frac{d^2}{d\zeta^2_2} + \frac{M_2\Omega^2_2}{2} \;  \zeta^2_2 \,\; \;\right]  \phi_{R \, m_1,\,m_2} (\zeta_1,\zeta_2) \; =   E_{m_1,\,m_2} \; \phi_{R\,  m_1,\,m_2} (\zeta_1,\zeta_2) \, , \label{schroed2} \end{equation}
\begin{equation} \left[ \,  \frac{p_1^2}{2 M_1}  - \frac{M_1(\hbar\,\Omega_1)^2}{2}   \frac{d^2}{dp^2_1}   + \frac{p_2^2}{2 M_2} - \frac{M_2(\hbar\,\Omega_2)^2}{2}   \frac{d^2}{dp^2_2}  \, \right]  \chi_{R \, m_1,\,m_2} (p_1,p_2)=  E_{m_1,\,m_2} \; \chi_{R\,  m_1,\,m_2} (p_1,p_2) \, . \label{fschroed2} \end{equation}
 In comparing Eqs.\ (\ref{schroed1}) and (\ref{fschroed1})  with Eqs.\ (\ref{schroed2}) and (\ref{fschroed2})  respectively  it gets clear that there hold the following identities:
\begin{equation} \zeta^2_1 \; = \; \frac{\hbar}{M_1\Omega_1} \; \rho^2_1 \; , \quad  \zeta^2_2 \; = \; \frac{\hbar}{M_2 \Omega_2} \; \rho^2_2\; , \quad p^2_1 \; = \; \hbar\,\Omega_1 M_1 \; \kappa^2_1 \; , \; p^2_2 \; = \; \hbar\,\Omega_2 M_2 \; \kappa^2_2  \; , \end{equation}
or, equivalently,
\begin{equation} \rho_1 \; = \; \left(\frac{M_1\Omega_1}{\hbar}\; \zeta^2_1 \right)^{\frac{1}{2}}  , \;  \rho_2 \; = \; \left(\frac{M_2 \Omega_2}{\hbar}\; \zeta^2_2\right)^{\frac{1}{2}}  , \; \kappa_1 \; = \; \left( \frac{p^2_1}{\hbar\,\Omega_1 M_1}\right)^{\frac{1}{2}} , \quad \kappa_2 \; = \; \left( \frac{p^2_2}{\hbar\,\Omega_2 M_2}\right)^{\frac{1}{2}} . \end{equation}
For the right eigen-functions $\phi_{R\,  m_1,\,m_2} (\zeta_1,\zeta_2)=\big<\!\big<\zeta_1,\,\zeta_2\,\big| m_1,\,m_2\big>$ and left eigen-functions $\phi_{L\,  m_1,\,m_2} (\zeta_1,\zeta_2)=\big<\!\big<m_1,\,m_2\,\big| \zeta_1,\,\zeta_2\big>$ of the Schr\"odinger Eq.\ (\ref{schroed2}) we have:
\begin{eqnarray}  \phi_{R\,  m_1,m_2} (\zeta_1,\,\zeta_2) & \propto & \phi_{m_1,m_2} \left(\left(\frac{M_1\Omega_1}{\hbar}\; \zeta^2_1 \right)^{\frac{1}{2}},\left(\frac{M_2\Omega_2}{\hbar}\; \zeta^2_2 \right)^{\frac{1}{2}}\right)  \; , \nonumber  \\[2mm]
\phi_{L\,  m_1,m_2} (\zeta_1,\,\zeta_2) & \propto & \phi^\ast_{m_1,m_2} \left(\left(\frac{M_1\Omega_1}{\hbar}\; \zeta^2_1 \right)^{\frac{1}{2}},\left(\frac{M_2\Omega_2}{\hbar}\; \zeta^2_2 \right)^{\frac{1}{2}}\right) \; . \label{wavfct1}
\end{eqnarray}
The analogous relations for the right eigen-functions $\chi_{R\,  m_1,\,m_2} (p_1,p_2)=\big<\!\big<p_1,\,p_2\,\big| m_1,\,m_2\big>$ and left eigen-functions $\chi_{L\,  m_1,\,m_2} (p_1,p_2)=\big<\!\big<m_1,\,m_2\,\big| p_1,\,p_2\big>$ of the Schr\"odinger Eq.\ (\ref{fschroed2}) are
\begin{eqnarray}  \chi_{R\,  m_1,m_2} (p_1,\,p_2) & \propto & \chi_{m_1,m_2} \left(\left( \frac{p^2_1}{\hbar\,\Omega_1 M_1}\right)^{\frac{1}{2}},\left( \frac{p^2_2}{\hbar\,\Omega_2 M_2}\right)^{\frac{1}{2}}\right)  \; , \nonumber  \\[2mm]
\chi_{L\,  m_1,m_2} (p_1,\,p_2) & \propto & \chi^\ast_{m_1,m_2} \left(\left( \frac{p^2_1}{\hbar\,\Omega_1 M_1}\right)^{\frac{1}{2}},\left( \frac{p^2_2}{\hbar\,\Omega_2 M_2}\right)^{\frac{1}{2}}\right) \; . \label{fwavfct1}
\end{eqnarray}
In standard Hermitian Quantum Mechanics the quantities $\hbar/(M_1\Omega_1)$ and $\hbar/(M_2\Omega_2)$ are considered to be the square of the oscillator length and are expected to be real-valued and positive. In the above mentioned case $M_1=-M_2$ one of these quantities has to be negative implying the respective oscillator length to be purely imaginary for a real-valued positive oscillator frequency. Hence, in order to maintain $\rho_1$ and $\rho_2$ to be real-valued, the variables $\zeta_1$ and $\zeta_2$ have to be chosen such that the expressions $M_1\,\Omega_1\,\zeta^2_1$ and $M_2\,\Omega_2\,\zeta^2_2$ are real-valued and non-negative. The analogous argument to maintain the momentum space variables $\kappa_1$ and $\kappa_2$ real-valued, the variables $p_1$ and $p_2$ should be chosen also such that the expressions $p_1^2/(M_1\Omega_1)$ and $p_2^2/(M_2\Omega_2)$ are real-valued and non-negative. The discussion shows that the configuration space variables  $\zeta_1$ and $\zeta_2$ and the momentum space variables  $p_1$ and $p_2$  should be considered --- for consistency reasons --- to be complex-valued and not merely real-valued.   Keeping this fact in mind we rewrite stationary Schr\"odinger Eq.\ (\ref{schroed2}) slightly as
\begin{equation} H_H(\zeta_1,\zeta_2)\; \phi_{R\,  m_1,\,m_2} (\zeta_1,\zeta_2) \; = \; E_{m_1,\,m_2} \; \phi_{R\,  m_1,\,m_2} (\zeta_1,\zeta_2) \label{schroed3}\; ,\end{equation}
with
\begin{eqnarray}
H_H(\zeta_1,\zeta_2) & = &  \frac{\hbar \, \Omega_1}{2} \left( \frac{M_1\Omega_1}{\hbar} \; \zeta^2_1 -\,\frac{\hbar}{M_1\Omega_1} \, \frac{d^2}{d\zeta^2_1} \right)    +  \,\frac{\hbar \,\Omega_2}{2} \left( \frac{M_2\Omega_2}{\hbar}\; \zeta^2_2 -\, \frac{\hbar}{M_2\Omega_2} \, \frac{d^2}{d\zeta^2_2} \right)  \; .\quad \end{eqnarray}
Recalling that there hold the identities

\begin{eqnarray} 
\lefteqn{ \frac{M_1\Omega_1}{\hbar} \; \zeta^2_1 -\,\frac{\hbar}{M_1\Omega_1} \, \frac{d^2}{d\zeta^2_1} \; =} \nonumber \\[2mm]
 & = & \left( \left(\frac{M_1\Omega_1}{2\hbar}\right)^{\frac{1}{2}}  \zeta_1 +\left(\frac{\hbar}{2M_1\Omega_1}\right)^{\frac{1}{2}}  \frac{d}{d\zeta_1} \right)\left( \left(\frac{M_1\Omega_1}{2\hbar}\right)^{\frac{1}{2}}  \zeta_1 -\,\left(\frac{\hbar}{2M_1\Omega_1}\right)^{\frac{1}{2}}  \frac{d}{d\zeta_1} \right)\nonumber \\
 & + & \left( \left(\frac{M_1\Omega_1}{2\hbar}\right)^{\frac{1}{2}}  \zeta_1 \,-\left(\frac{\hbar}{2M_1\Omega_1}\right)^{\frac{1}{2}}  \frac{d}{d\zeta_1} \right)\left( \left(\frac{M_1\Omega_1}{2\hbar}\right)^{\frac{1}{2}}  \zeta_1 +\,\left(\frac{\hbar}{2M_1\Omega_1}\right)^{\frac{1}{2}}  \frac{d}{d\zeta_1} \right)\; ,  \\[2mm]
\lefteqn{ \frac{M_2\Omega_2}{\hbar} \; \zeta^2_2 -\,\frac{\hbar}{M_2\Omega_2} \, \frac{d^2}{d\zeta^2_2} \; =} \nonumber \\[2mm]
 & = & \left( \left(\frac{M_2\Omega_2}{2\hbar}\right)^{\frac{1}{2}}  \zeta_2 \,+\left(\frac{\hbar}{2M_2\Omega_2}\right)^{\frac{1}{2}}  \frac{d}{d\zeta_2} \right)\left( \left(\frac{M_2\Omega_2}{2\hbar}\right)^{\frac{1}{2}}  \zeta_2 -\,\left(\frac{\hbar}{2M_2\Omega_2}\right)^{\frac{1}{2}}  \frac{d}{d\zeta_2} \right)\nonumber \\
 & + & \left( \left(\frac{M_2\Omega_2}{2\hbar}\right)^{\frac{1}{2}}  \zeta_2 \,-\left(\frac{\hbar}{2M_2\Omega_2}\right)^{\frac{1}{2}}  \frac{d}{d\zeta_2} \right)\left( \left(\frac{M_2\Omega_2}{2\hbar}\right)^{\frac{1}{2}}  \zeta_2 +\,\left(\frac{\hbar}{2M_2\Omega_2}\right)^{\frac{1}{2}}  \frac{d}{d\zeta_2} \right)\; ,  \end{eqnarray}
it is now straight forward to introduce annihilation and creation operators by performing the following identifications:

\begin{eqnarray} a_1\; & \leftrightarrow & \frac{1}{N_1}\left( \left(\frac{M_1\Omega_1}{2\hbar}\right)^{\frac{1}{2}}  \zeta_1 +\left(\frac{\hbar}{2M_1\Omega_1}\right)^{\frac{1}{2}}  \frac{d}{d\zeta_1} \right) \; , \nonumber \\[2mm]
c^+_1 & \leftrightarrow & N_1\,\left( \left(\frac{M_1\Omega_1}{2\hbar}\right)^{\frac{1}{2}}  \zeta_1 -\left(\frac{\hbar}{2M_1\Omega_1}\right)^{\frac{1}{2}}  \frac{d}{d\zeta_1} \right) \; , \nonumber \\[2mm]
 a_2\; & \leftrightarrow & \frac{1}{N_2}\left( \left( \frac{M_2\Omega_2}{2\hbar}\right)^{\frac{1}{2}}  \zeta_2 +\left(\frac{\hbar}{2M_2\Omega_2}\right)^{\frac{1}{2}}  \frac{d}{d\zeta_2} \right) \; , \nonumber \\[2mm]
 c^+_2 & \leftrightarrow & N_2\,\left( \left( \frac{M_2\Omega_2}{2\hbar}\right)^{\frac{1}{2}}  \zeta_2 -\left(\frac{\hbar}{2M_2\Omega_2}\right)^{\frac{1}{2}}  \frac{d}{d\zeta_2}  \right) \; , \label{creanop1}
\end{eqnarray}
with $N_1$ and $N_2$ being some arbitrary non-vanishing complex-valued constants. The creation and annihilation operators respect the following commutation relations
\begin{eqnarray} [\,a_1\, ,\, c^+_1\, ] & = & 1 \; , \quad  [\,a_1\, ,\, a_2\,\, ] \; = \; 0 \; , \quad  [\,a_1\, ,\, c^+_2\, ] \; = \; 0\, , 
\nonumber \\[2mm]  
[\,a_2\, ,\, c^+_2\, ] & = & 1   \; , \quad  [\,c^+_1\, ,\, c^+_2\, ] \; = \; 0 \; , \quad  [\,a_2\, ,\, c^+_1\, ] \; = \; 0 \; , \label{comrel1}\end{eqnarray}
holding even for arbitrary complex-valued mass parameters $M_1$ and $M_2$ and complex oscillator frequencies $\Omega_1$ and $\Omega_2$. 
Due to the fact that the configuration space variables $\zeta_1$ and $\zeta_2$ should be considered complex-valued it turns out that the annihilation operators $a_1$ or $a_2$ and respective annihilation operators $c_1^+$ or $c^+_2$ are not related by a naive Hermitian conjugation.

 Keeping in mind  the considerations above there obviously holds the following identification on the level of Hamilton operators:
\begin{eqnarray} H_H & = & \frac{\hbar\, \Omega_1}{2}\left( c_1^+  a_1 + a_1 \,c^+_1\right) \; + \; \frac{\hbar\, \Omega_2}{2}\left( c_2^+ a_2 + a_2 \,c^+_2\right) \quad \leftrightarrow \quad H_H(\zeta_1,\zeta_2) \; . \label{xyham1}
\end{eqnarray}
The  here so-called holomorphic Hamilton operator $ H_H(\zeta_1,\zeta_2)$ acts on wave functions $\psi_H(\zeta_1,\zeta_2)$ being holomorphic functions in the configuration space variables $\zeta_1$ and $\zeta_2$.
Already a naive Hermitian conjugation on the level of operators yielding the well known identity $(AB)^+=B^+A^+$ shows that the Hamilton operator $H_H$ cannot be Hermitian even for real-valued oscillator frequencies as there holds $H_A=H_H^+ \not= H_H$, i.\ e.,
\begin{eqnarray} H^+_H & = & \frac{\hbar\, \Omega^\ast_1}{2}\left( a_1^+  c_1 + c_1 \,a^+_1\right) \; + \; \frac{\hbar\, \Omega^\ast_2}{2}\left( a_2^+ c_2 + c_2 \,a^+_2\right) \; = \; H_A \quad \leftrightarrow \quad H_A(\zeta^\ast_1,\zeta^\ast_2) \; , \label{xyham2}
\end{eqnarray}
with --- after partially integrating  differential operators acting originally to the left ---
\begin{eqnarray}
H_A(\zeta^\ast_1,\zeta^\ast_2) & = &  \frac{\hbar \, \Omega^\ast_1}{2} \left( \frac{M^\ast_1\Omega^\ast_1}{\hbar} \; \zeta^{\ast2}_1 -\,\frac{\hbar}{M^\ast_1\Omega^\ast_1} \, \frac{d^2}{d\zeta^{\ast2}_1} \right)    +  \,\frac{\hbar \,\Omega^\ast_2}{2} \left( \frac{M^\ast_2\Omega^\ast_2}{\hbar}\; \zeta^{\ast 2}_2 -\, \frac{\hbar}{M^\ast_2\Omega^\ast_2} \, \frac{d^2}{d\zeta^{\ast 2}_2} \right)  \; .\quad \end{eqnarray}
 The here so called anti-holomorphic Hamilton operator $H_A(\zeta^\ast_1,\zeta^\ast_2)$ acting  on wave functions $\psi_A(\zeta^\ast_1,\zeta^\ast_2)$  being anti-holomorphic functions in the complex-conjugate configuration space variables $\zeta^\ast_1$ and $\zeta^\ast_2$ is expressed in terms of a set of creation operators and annihilation operators obtained by Hermitian conjugation of the set underlying the holomorphic Hamilton operator $H_H$, i.\ e.,
\begin{eqnarray} a^+_1 & \leftrightarrow & \frac{1}{N^\ast_1}\left( \left(\frac{M^\ast_1\Omega^\ast_1}{2\hbar}\right)^{\frac{1}{2}}  \zeta^\ast_1 -\left(\frac{\hbar}{2M^\ast_1\Omega^\ast_1}\right)^{\frac{1}{2}}  \frac{d}{d\zeta^\ast_1} \right) \; , \nonumber \\[2mm]
c_1 \; & \leftrightarrow & N^\ast_1\,\left( \left(\frac{M^\ast_1\Omega^\ast_1}{2\hbar}\right)^{\frac{1}{2}}  \zeta^\ast_1 +\left(\frac{\hbar}{2M^\ast_1\Omega^\ast_1}\right)^{\frac{1}{2}}  \frac{d}{d\zeta^\ast_1} \right) \; , \nonumber \\[2mm]
 a^+_2  & \leftrightarrow & \frac{1}{N^\ast_2}\left( \left( \frac{M^\ast_2\Omega^\ast_2}{2\hbar}\right)^{\frac{1}{2}}  \zeta^\ast_2 -\left(\frac{\hbar}{2M^\ast_2\Omega^\ast_2}\right)^{\frac{1}{2}}  \frac{d}{d\zeta^\ast_2} \right) \; , \nonumber  \\[2mm]
 c_2 \; & \leftrightarrow & N^\ast_2\,\left( \left( \frac{M^\ast_2\Omega^\ast_2}{2\hbar}\right)^{\frac{1}{2}}  \zeta^\ast_2 +\left(\frac{\hbar}{2M^\ast_2\Omega^\ast_2}\right)^{\frac{1}{2}}  \frac{d}{d\zeta^\ast_2}  \right) \; . \label{creanop2}
\end{eqnarray}
They commute with all the creation and annihilation operators underlying the holomorphic Hamilton operator $H_H$  (implying also $[H_H,H_A]=0$) and respect furthermore the following commutation relations obtained by Hermitian conjugation of Eq.\ (\ref{comrel1}):
\begin{eqnarray} [\,c_1\, ,\, a^+_1\, ] & = & 1 \; , \quad  [\,c_1\,\, ,\,\, c_2\,\, ] \; = \; 0 \; , \quad  [\,c_1\, ,\, a^+_2\, ] \; = \; 0\, , 
\nonumber \\[2mm]  
[\,c_2\, ,\, a^+_2\, ] & = & 1   \; , \quad  [\,a^+_1\, ,\, a^+_2\, ] \; = \; 0 \; , \quad  [\,c_2\, ,\, a^+_1\, ] \; = \; 0 \;. \label{comrel2} \end{eqnarray}
It is straight forward to construct the spectrum of $H_A$ by performing a complex conjugation of the spectrum of $H_H$ found in Eq.\ (\ref{spec1}), i.\ e.,
\begin{equation} E^{\,\ast}_{n_1,\,n_2} \; = \; \hbar \, \Omega^\ast_1 \left( n_1 + \frac{1}{2}\right) +  \hbar \,\Omega^\ast_2 \left( n_2 + \frac{1}{2}\right) \label{spec2} \end{equation}
with $n_1$ and $n_2$ being arbitrary non-negative integers. By adding to the (non-Hermitian) homomorphic Hamilton operator $H_H$ the (also non-Hermitian) anti-holomorphic Hamilton operator $H_A$ we perform a here so-called $PT$-symmetric completion and  obtain a $PT$-symmetric \cite{Bender:2007nj}\cite{Bender:2004sv}\cite{Bender:1998ke}\cite{Weigert:2003py}\cite{Znojil:2001xy} Hamilton operator $H = H_H + H_A$ possessing the following spectrum:
\begin{eqnarray}  \lefteqn{E_{m_1,\,m_2\,;\,n_1,\,n_2} \; = \; E_{m_1,\,m_2} + E^{\,\ast}_{n_1,\,n_2} \; =} \nonumber \\[3mm]
 & = & \hbar \, \Omega_1 \left( m_1 + \frac{1}{2}\right) +  \hbar \,\Omega_2 \left( m_2 + \frac{1}{2}\right)  \;+ \;\hbar \, \Omega^\ast_1 \left( n_1 + \frac{1}{2}\right) +  \hbar \,\Omega^\ast_2 \left( n_2 + \frac{1}{2}\right) \; , \label{spec3} \end{eqnarray}
with $m_1$, $m_2$, $n_1$ and $n_2$ being arbitrary non-negative integers. The underlying $PT$-symmetry of the Hamilton operator $H=H_H + H_A$ being expressed here by the relation $H^+_H=H_A$ leads to the fact that the eigen-values of the Hamilton operator are either real-valued or appear in complex-conjugate pairs. In our notation this property is represented by the following identity:
\begin{equation} E^\ast_{m_1,\,m_2\,;\,n_1,\,n_2} \; = \; E_{n_1,\,n_2\,;\,m_1,\,m_2} \; .\end{equation}
The construction of the bi-orthonormal and complete system of right eigen-states $\left|m_1,\,m_2\,;\,n_1,\,n_2\right>$ and left eigen-states $\left<\!\left<m_1,\,m_2\,;\,n_1,\,n_2\right|\right.$ of the Hamilton operator $H=H_H+H_A$ respecting the stationary Schr\"odinger equations
\begin{eqnarray} H \, \left|m_1,\,m_2\,;\,n_1,\,n_2\right> & = & E_{m_1,\,m_2\,;\,n_1,\,n_2} \, \left|m_1,\,m_2\,;\,n_1,\,n_2\right> \; , \nonumber \\[2mm]
 \left<\!\left<m_1,\,m_2\,;\,n_1,\,n_2\right|\right. H & = & E_{m_1,\,m_2\,;\,n_1,\,n_2}\, \left<\!\left<m_1,\,m_2\,;\,n_1,\,n_2\right|\right. \; , \end{eqnarray} 
going back to ideas of N.\ Nakanishi \cite{Nakanishi:1972wx}\cite{Nakanishi:1972pt} and M.\ Froissart \cite{Froissart:1959occ} has been discussed e.\ g.\ in \cite{Kleefeld:2003zj}\cite{Kleefeld:2004jb}\cite{Kleefeld:2004qs} (see also \cite{Kleefeld:2002au}\cite{Kleefeld:2002gw}).  In the standard manner we introduce a right vacuum state $\big|0\big>\stackrel{!}{=}\big|0\big>\!\big>=\big<\!\big<0\big|^+$ and a left vacuum state $\big<\!\big<0\big|\stackrel{!}{=}\big<0\big|=\big|0\big>^+$ annihilating respectively annihilation operators or creation operators:
\begin{equation} a_1 \;\big|0\big> \; = \; a_2 \;\big|0\big> \; = \; c_1 \;\big|0\big> \; = \; c_2 \;\big|0\big>\; = \; 0 \; ,  \;  \big<\!\big<0\big| \; a_1^+ \; = \; \big<\!\big<0\big| \; a_2^+ \; = \; \big<\!\big<0\big| \; c_1^+ \; = \; \big<\!\big<0\big| \; c_2^+ \; = \; 0 \; .\end{equation}
The left and right vacuum states are chosen to respect the normalization condition $\big<\!\big<0\big|0\big>\; = \; 1$. In analogy to standard Hermitian Quantum Theory it is now straight forward to denote the  right eigen-states $\left|m_1,\,m_2\,;\,n_1,\,n_2\right>$ and left eigen-states $\left<\!\left<m_1,\,m_2\,;\,n_1,\,n_2\right|\right.$ of the Hamilton operator $H=H_H+H_A$ in the following way:
\begin{eqnarray} \left|m_1,m_2\,;\,n_1,n_2\right> & \equiv &  \frac{1}{N_{m_1,\,m_2\,;\,n_1,\,n_2}\;\sqrt{m_1!\,m_2!\,n_1!\,n_2!}} \;  (c_1^+)^{m_1} \, (c_2^+)^{m_2} \, (a_1^+)^{n_1} \, (a_2^+)^{n_2}  \left|0\right> \, , \quad \label{eigen1} \\[2mm]
\left<\!\left<m_1,m_2\,;\,n_1,n_2\right|\right. & \equiv &  \frac{N_{m_1,\,m_2\,;\,n_1,\,n_2}}{\sqrt{m_1!\,m_2!\,n_1!\,n_2!}} \; \left<\!\left<0\right|\right. c^{n_2} \, c^{n_1} \, a^{m_2} \, a^{m_1} \; , \label{eigen2} 
\end{eqnarray}
with $N_{m_1,\,m_2\,;\,n_1,\,n_2}$ being some arbitrary non-vanishing eventually complex-valued constants. The orthonormality relations for these states are generalized in the following way to the square-integrable left and right eigen-functions of the PT-symmetric Hamilton operator $H=H_H + H_A$:
\begin{eqnarray}  \lefteqn{\delta_{m_1,\,m_1^\prime}  \; \delta_{m_2,\,m_2^\prime}\;  \delta_{n_1,\,n_1^\prime}  \; \delta_{n_2,\,n_2^\prime} \; = \; \big<\!\big<m_1,m_2\,;\,n_1,n_2\big|m^\prime_1,m^\prime_2\,;\,n^\prime_1,n^\prime_2\big> \; =} \nonumber \\[2mm]
 & = & \int d\zeta_1  \int d\zeta_2 \int d\zeta^\ast_1 \! \int d\zeta^\ast_2  \;\; \phi_{L\, m_1,m_2\,;\,n_1,n_2}(\zeta_1,\zeta_2\,;\,\zeta^\ast_1,\zeta^\ast_2) \;\; \phi_{R\, m^\prime_1,m^\prime_2\,;\,n^\prime_1,n^\prime_2}(\zeta_1,\zeta_2\,;\,\zeta^\ast_1,\zeta^\ast_2)\nonumber \\
 & = & \int d\zeta_1  \int d\zeta_2 \int d\zeta^\ast_1 \! \int d\zeta^\ast_2  \; \big<\!\big<m_1,m_2\,;\,n_1,n_2\big|\zeta_1,\zeta_2\,;\,\zeta^\ast_1,\zeta^\ast_2\big>\big<\!\big<\zeta_1,\zeta_2\,;\,\zeta^\ast_1,\zeta^\ast_2\big|m^\prime_1,m^\prime_2\,;\,n^\prime_1,n^\prime_2\big> \nonumber \\
 & = & \int d\zeta_1  \int d\zeta_2 \int d\zeta^\ast_1 \! \int d\zeta^\ast_2  \; \big<\!\big<m_1,m_2\big|\zeta_1,\zeta_2\big>\big<\!\big<\zeta_1,\zeta_2\big|n_1,n_2\big>^\ast \big<\!\big<\zeta_1,\zeta_2\big|m^\prime_1,m^\prime_2\big>\big<\!\big<n^\prime_1,n^\prime_2\big|\zeta_1,\zeta_2\big>^\ast \nonumber \\
 & = & \int d\zeta_1  \int d\zeta_2 \int d\zeta^\ast_1 \! \int d\zeta^\ast_2  \; \big<\!\big<m_1,m_2\big|\zeta_1,\zeta_2\big>\big<n_1,n_2\big|\zeta^\ast_1,\zeta^\ast_2\big>\!\big> \; \big<\!\big<\zeta_1,\zeta_2\big|m^\prime_1,m^\prime_2\big>\big<\zeta^\ast_1,\zeta^\ast_2\big|n^\prime_1,n^\prime_2\big>\!\big> \; , \nonumber \\
 & = & \int \!\frac{dp_1}{2\pi \hbar}  \int \!\frac{dp_2}{2\pi \hbar} \int \!\frac{dp^\ast_1}{2\pi \hbar} \! \int \!\frac{dp^\ast_2}{2\pi \hbar}  \;\; \chi_{L\, m_1,m_2\,;\,n_1,n_2}(p_1,p_2\,;\,p^\ast_1,p^\ast_2) \;\; \chi_{R\, m^\prime_1,m^\prime_2\,;\,n^\prime_1,n^\prime_2}(p_1,p_2\,;\,p^\ast_1,p^\ast_2)\nonumber \\
 & = &  \int \!\frac{dp_1}{2\pi \hbar}  \int \!\frac{dp_2}{2\pi \hbar} \int \!\frac{dp^\ast_1}{2\pi \hbar} \! \int \!\frac{dp^\ast_2}{2\pi \hbar}   \; \big<\!\big<m_1,m_2\,;\,n_1,n_2\big|p_1,p_2\,;\,p^\ast_1,p^\ast_2\big>\big<\!\big<p_1,p_2\,;\,p^\ast_1,p^\ast_2\big|m^\prime_1,m^\prime_2\,;\,n^\prime_1,n^\prime_2\big> \nonumber \\
 & = &  \int \!\frac{dp_1}{2\pi \hbar}  \int \!\frac{dp_2}{2\pi \hbar} \int \!\frac{dp^\ast_1}{2\pi \hbar} \! \int \!\frac{dp^\ast_2}{2\pi \hbar}  \; \big<\!\big<m_1,m_2\big|p_1,p_2\big>\big<\!\big<p_1,p_2\big|n_1,n_2\big>^\ast \big<\!\big<p_1,p_2\big|m^\prime_1,m^\prime_2\big>\big<\!\big<n^\prime_1,n^\prime_2\big|p_1,p_2\big>^\ast \nonumber \\
 & = &  \int \!\frac{dp_1}{2\pi \hbar}  \int \!\frac{dp_2}{2\pi \hbar} \int \!\frac{dp^\ast_1}{2\pi \hbar} \! \int \!\frac{dp^\ast_2}{2\pi \hbar}   \; \big<\!\big<m_1,m_2\big|p_1,p_2\big>\big<n_1,n_2\big|p^\ast_1,p^\ast_2\big>\!\big> \; \big<\!\big<p_1,p_2\big|m^\prime_1,m^\prime_2\big>\big<p^\ast_1,p^\ast_2\big|n^\prime_1,n^\prime_2\big>\!\big> \; . \nonumber \\
\end{eqnarray}
As pointed out in the context of Eq.\ (\ref{wavfct1}) the integration contours for $\zeta_1$ and $\zeta_2$ have to be chosen such that respectively the expressions $M_1\Omega_1\zeta_1^2$ and $M_2\Omega_2\zeta_2^2$ are real-valued and non-negative. Analogously the integration contours for $p_1$ and $p_2$ have to be chosen such that respectively the expressions $p^2_1/(M_1\Omega_1)$ and $p^2_2/(M_2\Omega_2)$ are real-valued and non-negative.   The integration contours for $\zeta^\ast_1$, $\zeta^\ast_2$,  $p^\ast_1$ and $p^\ast_2$ are obtained by complex conjugation of the respective integration contours for $\zeta_1$, $\zeta_2$, $p_1$ and $p_2$. 

The square integrability of the eigen-functions underlines that there exists a Fourier-transform interrelating $\phi_{R\, m_1,m_2\,;\,n_1,n_2}(\zeta_1,\zeta_2\,;\,\zeta^\ast_1,\zeta^\ast_2)$ and $\phi_{L\, m_1,m_2\,;\,n_1,n_2}(\zeta_1,\zeta_2\,;\,\zeta^\ast_1,\zeta^\ast_2)$ with respectively $\chi_{R\, m_1,m_2\,;\,n_1,n_2}(p_1,p_2\,;\,p^\ast_1,p^\ast_2)$ and $\chi_{L\, m^\prime_1,m^\prime_2\,;\,n^\prime_1,n^\prime_2}(p_1,p_2\,;\,p^\ast_1,p^\ast_2)$ along the complex-valued integration contours along which they are square-integrable. 
More explicitly there holds
\begin{eqnarray} \lefteqn{\phi_{R\, m_1,m_2\,;\,n_1,n_2}(\zeta_1,\zeta_2\,;\,\zeta^\ast_1,\zeta^\ast_2) \; =} \nonumber \\[2mm]
 & = &  \int \!\frac{dp_1}{2\pi \hbar}  \int \!\frac{dp_2}{2\pi \hbar} \int \!\frac{dp^\ast_1}{2\pi \hbar} \! \int \!\frac{dp^\ast_2}{2\pi \hbar}  \;\; e^{+ \frac{i}{\hbar} (p_1 \zeta_1 + p_2 \, \zeta_2+p^\ast_1 \zeta^\ast_1 + p^\ast_2 \, \zeta^\ast_2)} \; \chi_{R\, m_1,m_2\,;\,n_1,n_2}(p_1,p_2\,;\,p^\ast_1,p^\ast_2) \; , \nonumber \\[2mm]
 \lefteqn{\phi_{L\, m_1,m_2\,;\,n_1,n_2}(\zeta_1,\zeta_2\,;\,\zeta^\ast_1,\zeta^\ast_2) \; =} \nonumber \\[2mm]
 & = &  \int \!\frac{dp_1}{2\pi \hbar}  \int \!\frac{dp_2}{2\pi \hbar} \int \!\frac{dp^\ast_1}{2\pi \hbar} \! \int \!\frac{dp^\ast_2}{2\pi \hbar}  \;\; e^{- \frac{i}{\hbar} (p_1 \zeta_1 + p_2 \, \zeta_2+p^\ast_1 \zeta^\ast_1 + p^\ast_2 \, \zeta^\ast_2)} \; \chi_{L\, m_1,m_2\,;\,n_1,n_2}(p_1,p_2\,;\,p^\ast_1,p^\ast_2) \; , \nonumber \\[2mm]
\lefteqn{\chi_{R\, m_1,m_2\,;\,n_1,n_2}(p_1,p_2\,;\,p^\ast_1,p^\ast_2)   \; =} \nonumber \\[2mm]
 & = &  \int d\zeta_1  \int d\zeta_2 \int d\zeta^\ast_1 \! \int d\zeta^\ast_2   \;\; e^{- \frac{i}{\hbar} (p_1 \zeta_1 + p_2 \, \zeta_2+p^\ast_1 \zeta^\ast_1 + p^\ast_2 \, \zeta^\ast_2)} \; \phi_{R\, m_1,m_2\,;\,n_1,n_2}(\zeta_1,\zeta_2\,;\,\zeta^\ast_1,\zeta^\ast_2) \; , \nonumber \\[2mm]
\lefteqn{\chi_{L\, m_1,m_2\,;\,n_1,n_2}(p_1,p_2\,;\,p^\ast_1,p^\ast_2)   \; =} \nonumber \\[2mm]
 & = &  \int d\zeta_1  \int d\zeta_2 \int d\zeta^\ast_1 \! \int d\zeta^\ast_2   \;\; e^{+ \frac{i}{\hbar} (p_1 \zeta_1 + p_2 \, \zeta_2+p^\ast_1 \zeta^\ast_1 + p^\ast_2 \, \zeta^\ast_2)} \; \phi_{L\, m_1,m_2\,;\,n_1,n_2}(\zeta_1,\zeta_2\,;\,\zeta^\ast_1,\zeta^\ast_2) \; , 
\end{eqnarray}
or, equivalently, in (non-Hermitian) ``bra-ket" \cite{Kleefeld:2003zj} notation:
\begin{eqnarray} \lefteqn{\big<\!\big<\zeta_1,\zeta_2\,;\,\zeta^\ast_1,\zeta^\ast_2\big|m_1,m_2\,;\,n_1,n_2\big> \; =} \nonumber \\[2mm]
 & = &  \int \!\frac{dp_1}{2\pi \hbar}  \int \!\frac{dp_2}{2\pi \hbar} \int \!\frac{dp^\ast_1}{2\pi \hbar} \! \int \!\frac{dp^\ast_2}{2\pi \hbar}  \;\; \big<\!\big<\zeta_1,\zeta_2\,;\,\zeta^\ast_1,\zeta^\ast_2\big|p_1,p_2\,;\,p^\ast_1,p^\ast_2\big>  \big<\!\big<p_1,p_2\,;\,p^\ast_1,p^\ast_2\big|m_1,m_2\,;\,n_1,n_2\big> \; , \nonumber \\[2mm]
\lefteqn{\big<\!\big<m_1,m_2\,;\,n_1,n_2\big|\zeta_1,\zeta_2\,;\,\zeta^\ast_1,\zeta^\ast_2\big> \; =} \nonumber \\[2mm]
 & = &  \int \!\frac{dp_1}{2\pi \hbar}  \int \!\frac{dp_2}{2\pi \hbar} \int \!\frac{dp^\ast_1}{2\pi \hbar} \! \int \!\frac{dp^\ast_2}{2\pi \hbar}  \;\; \big<\!\big<m_1,m_2\,;\,n_1,n_2\big|p_1,p_2\,;\,p^\ast_1,p^\ast_2\big>\big<\!\big<p_1,p_2\,;\,p^\ast_1,p^\ast_2\big|\zeta_1,\zeta_2\,;\,\zeta^\ast_1,\zeta^\ast_2\big> \; , \nonumber \\[2mm]
 \lefteqn{\big<\!\big<p_1,p_2\,;\,p^\ast_1,p^\ast_2\big|m_1,m_2\,;\,n_1,n_2\big> \; =} \nonumber \\[2mm]
 & = &  \int d\zeta_1  \int d\zeta_2 \int d\zeta^\ast_1 \! \int d\zeta^\ast_2 \;\; \big<\!\big<p_1,p_2\,;\,p^\ast_1,p^\ast_2\big|\zeta_1,\zeta_2\,;\,\zeta^\ast_1,\zeta^\ast_2\big>  \big<\!\big<\zeta_1,\zeta_2\,;\,\zeta^\ast_1,\zeta^\ast_2\big|m_1,m_2\,;\,n_1,n_2\big> \; , \nonumber \\[2mm]
\lefteqn{\big<\!\big<m_1,m_2\,;\,n_1,n_2\big|p_1,p_2\,;\,p^\ast_1,p^\ast_2\big> \; =} \nonumber \\[2mm]
 & = &  \int d\zeta_1  \int d\zeta_2 \int d\zeta^\ast_1 \! \int d\zeta^\ast_2  \;\; \big<\!\big<m_1,m_2\,;\,n_1,n_2\big|\zeta_1,\zeta_2\,;\,\zeta^\ast_1,\zeta^\ast_2\big>\big<\!\big<\zeta_1,\zeta_2\,;\,\zeta^\ast_1,\zeta^\ast_2\big|p_1,p_2\,;\,p^\ast_1,p^\ast_2\big> \; .
\end{eqnarray}
The discussion shows that there holds the completeness relation
\begin{eqnarray}  I & = & \int d\zeta_1  \int d\zeta_2 \int d\zeta^\ast_1 \! \int d\zeta^\ast_2  \;\;\big|\zeta_1,\zeta_2\,;\,\zeta^\ast_1,\zeta^\ast_2\big>\big<\!\big<\zeta_1,\zeta_2\,;\,\zeta^\ast_1,\zeta^\ast_2\big| \nonumber \\
 & = & \int \!\frac{dp_1}{2\pi \hbar}  \int \!\frac{dp_2}{2\pi \hbar} \int \!\frac{dp^\ast_1}{2\pi \hbar} \! \int \!\frac{dp^\ast_2}{2\pi \hbar}  \;\; \big|p_1,p_2\,;\,p^\ast_1,p^\ast_2\big>  \big<\!\big<p_1,p_2\,;\,p^\ast_1,p^\ast_2\big| \; ,
\end{eqnarray}
with $I$ being the identity operator, while the matrix elements 
\begin{eqnarray} \big<\!\big<\zeta_1,\zeta_2\,;\,\zeta^\ast_1,\zeta^\ast_2\big|\zeta^\prime_1,\zeta^\prime_2\,;\,\zeta^{\ast\prime}_1,\zeta^{\ast\prime}_2\big>\, & = & \delta (\zeta_1 - \zeta^\prime_1) \; \delta (\zeta_2- \zeta^\prime_2) \; \delta (\zeta^\ast_1 - \zeta^{\ast \prime}_1) \; \delta (\zeta^\ast_2 - \zeta^{\ast \prime}_2) \; , \nonumber \\[2mm] 
 \big<\!\big<p_1,p_2\,;\,p^\ast_1,p^\ast_2\big|p^\prime_1,p^\prime_2\,;\,p^{\ast\prime}_1,p^{\ast\prime}_2\big> & = & 2\pi\hbar\, \delta (p_1 - p^\prime_1) \; 2\pi\hbar\, \delta (p_2- p^\prime_2) \;2\pi\hbar\,  \delta (p^\ast_1 - p^{\ast \prime}_1) \; 2\pi\hbar\, \delta (p^\ast_2 - p^{\ast \prime}_2) \; , \nonumber \\
\end{eqnarray}
should be interpreted as $\delta$-distributions with complex-valued arguments along the complex-valued integration contours.

 For convenience of the reader we want to provide here also the appropriately orthonormalized right and left ground-state wave functions $ \phi_{R\; 0,0\,;\,0,0}(\zeta_1,\zeta_2\,;\,\zeta^\ast_1,\zeta^\ast_2) $ and $ \phi_{L\; 0,0\,;\,0,0}(\zeta_1,\zeta_2\,;\,\zeta^\ast_1,\zeta^\ast_2) $ respectively obtained by solving the stationary Schr\"odinger equation for $H(\zeta_1,\zeta_2;\zeta^\ast_1,\zeta^\ast_2)=H_H(\zeta_1,\zeta_2)+H_A(\zeta^\ast_1,\zeta^\ast_2)$ and the energy eigen-value $E_{\,0,\,0\,;\,0,\,0} = \frac{1}{2} ( \hbar \Omega_1 + \hbar \Omega_2 + \hbar\Omega^\ast_1+\hbar\Omega^\ast_2)\;$:

\begin{eqnarray}  \lefteqn{\phi_{R\; 0,0\,;\,0,0}(\zeta_1,\zeta_2\,;\,\zeta^\ast_1,\zeta^\ast_2) \; = \; \big<\!\big<\zeta_1,\zeta_2\,;\,\zeta^\ast_1,\zeta^\ast_2\big|0,0;0,0\big> \; = \; \big<\!\big<\zeta_1,\zeta_2\,;\,\zeta^\ast_1,\zeta^\ast_2\big|0\big> \; =} \nonumber \\[2mm]
 \lefteqn{\phi_{L\; 0,0\,;\,0,0}(\zeta_1,\zeta_2\,;\,\zeta^\ast_1,\zeta^\ast_2) \; = \; \big<\!\big<0,0;0,0\big|\zeta_1,\zeta_2\,;\,\zeta^\ast_1,\zeta^\ast_2\big> \; = \; \big<\!\big<0\big|\zeta_1,\zeta_2\,;\,\zeta^\ast_1,\zeta^\ast_2\big> \; =} \nonumber \\[1mm]
 & = & \frac{1}{\pi \;\sqrt{\left| \frac{\hbar}{M_1\Omega_1}\right|\,\left|\frac{\hbar}{M_2\Omega_2} \right|}} \; \exp\left( - \,\frac{M_1\Omega_1\,\zeta_1^2 +M_2\Omega_2\,\zeta_2^2 +M^\ast_1\Omega^\ast_1\,\zeta_1^{\ast\,2} +M^\ast_2\Omega^\ast_2\,\zeta_2^{\ast\,2}}{2\,\hbar} \right) \; .
\end{eqnarray}
The wave functions of further eigen-states are obtained by applying the differential operators of Eq.\ (\ref{creanop1}) and (\ref{creanop2}) to the vacuum as specified in Eq.\ (\ref{eigen1}) and Eq.\ (\ref{eigen2}).
\section{Canonical quantization of two independent Bosonic Harmonic Oscillators}
Here we want to show that the results obtained in the previous section can be obtained in a straight forward way also by canonical quantization of the PT-symmetry completed system of two independent Bosonic Harmonic Oscillators. Starting point is the holomorphic Lagrange function provided in Eq.\ (\ref{lagtwoho1}), i.\ e.,
\begin{equation} L_H[Q_1,Q_2,\dot{Q}_1,\dot{Q}_2] \; = \; \frac{M_1}{2} \left( \dot{Q}_{1} (t)^2 \, - \, \Omega^2_1  \, Q_1(t)^2\right) \; + \; \frac{M_2}{2} \left( \dot{Q}_{2} (t)^2 \, - \, \Omega^2_2  \, Q_2(t)^2\right) \; . \label{lagr1}
\end{equation}
The corresponding anti-holomorphic Lagrange function is given by
\begin{equation} L_A[Q^+_1,Q^+_2,\dot{Q}^+_1,\dot{Q}^+_2] \; = \; \frac{M^\ast_1}{2} \left( \dot{Q}_{1} (t)^{+\,2} \, - \, \Omega^{\ast\,2}_1  \, Q_1(t)^{+\,2}\right) \; + \; \frac{M^\ast_2}{2} \left( \dot{Q}_{2} (t)^{+\,2} \, - \, \Omega^{\ast\,2}_2  \, Q_2(t)^{+\,2}\right) \; . \label{lagr2}
\end{equation}
Here $Q^+_1$ and $Q^+_2$ should be considered as two new independent degrees of freedom. The PT-symmetric completion of the system is performed by constructing a new Lagrange function by adding the respective holomorphic and anti-holomorphic Lagrange functions, i.\ e.,

\begin{eqnarray} \lefteqn{ L[Q_1,Q_2,\dot{Q}_1,\dot{Q}_2,Q^+_1,Q^+_2,\dot{Q}^+_1,\dot{Q}^+_2] \; = } \nonumber \\[2mm] 
 & = &  L_H[Q_1,Q_2,\dot{Q}_1,\dot{Q}_2] +  L_A[Q^+_1,Q^+_2,\dot{Q}^+_1,\dot{Q}^+_2]  \nonumber \\[2mm]
 & = & \; \frac{M_1}{2} \left( \dot{Q}_{1} (t)^2 \, - \, \Omega^2_1  \, Q_1(t)^2\right)\quad\;\;\; \; \; + \; \frac{M_2}{2} \left( \dot{Q}_{2} (t)^2 \, - \, \Omega^2_2  \, Q_2(t)^2\right) \nonumber \\
 & + & \; \frac{M^\ast_1}{2} \left( \dot{Q}_{1} (t)^{+\,2} \, - \, \Omega^{\ast\,2}_1  \, Q_1(t)^{+\,2}\right) \; + \; \frac{M^\ast_2}{2} \left( \dot{Q}_{2} (t)^{+\,2} \, - \, \Omega^{\ast\,2}_2  \, Q_2(t)^{+\,2}\right) \; . \label{lagr3}
\end{eqnarray}
Canonical conjugate momenta are derived in the standard manner:

\begin{eqnarray} P_1(t)\;\; & = & \frac{\delta L}{\delta \dot{Q}_1(t)} \;\;\; = \; M_1 \, \dot{Q}_1 (t) \; , \qquad   P_2(t) \;\;\; = \; \frac{\delta L}{\delta \dot{Q}_2(t)} \;\;\;\; = \; M_2 \, \dot{Q}_2 (t) \; , \label{xmom1} \\[2mm]
 P_1(t)^+ & = & \frac{\delta L}{\delta \dot{Q}_1(t)^+} \; = \; M^\ast_1 \, \dot{Q}_1 (t)^+ \; ,  \quad\;\,  P_2(t)^+ \; = \; \frac{\delta L}{\delta \dot{Q}_2(t)^+} \; = \; M^\ast_2 \, \dot{Q}_2 (t)^+ \; . \label{xmom2}
\end{eqnarray}
The Lagrange equations of motion are obviously:

\begin{eqnarray} \frac{d}{dt} \,  \frac{\delta L}{\delta \dot{Q}_1(t)} \, - \, \frac{\delta L}{\delta Q_1(t)} \; = \; 0  & \Rightarrow & M_1 \left(  \ddot{Q}_1(t) \, + \, \Omega^2_1 \, Q_1(t) \right) \; = \; 0 \; , \\[2mm]
\frac{d}{dt} \,  \frac{\delta L}{\delta \dot{Q}_2(t)} \, - \, \frac{\delta L}{\delta Q_2(t)} \; = \; 0 & \Rightarrow & M_2 \left(  \ddot{Q}_2(t) \, + \, \Omega^2_2 \, Q_2(t) \right) \; = \; 0 \; , \\[2mm]
 \frac{d}{dt} \,  \frac{\delta L}{\delta \dot{Q}_1(t)^+} \, - \, \frac{\delta L}{\delta Q_1(t)^+} \; = \; 0  & \Rightarrow & M^\ast_1 \left(  \ddot{Q}_1(t)^+ \, + \, \Omega^{\ast\, 2}_1 \, Q_1(t)^+ \right) \; = \; 0 \; , \\[2mm]
 \frac{d}{dt} \,  \frac{\delta L}{\delta \dot{Q}_2(t)^+} \, - \, \frac{\delta L}{\delta Q_2(t)^+} \; = \; 0  & \Rightarrow & M^\ast_2 \left(  \ddot{Q}_2(t)^+ \, + \, \Omega^{\ast\, 2}_2 \, Q_2(t)^+ \right) \; = \; 0 \; .
\end{eqnarray}
The solutions of these equations have been given for $Q_1$ and $Q_2$ in Eq.\ (\ref{qsolut1}). Considering also $Q^+_1$ and $Q^+_2$ we have:
\begin{eqnarray} Q_{1} (t)\;\;\; & = & Q^{(+)}_1 (t) \; \;\; + \; Q^{(-)}_1 (t) \quad  = \;   Q^{(+)}_1 (0) \; e^{-\,i\,\Omega_1 t}\;\;\;\, + \; Q^{(-)}_1 (0) \; e^{+\,i\,\Omega_1 t} \; , \nonumber \\[2mm]
 Q_{2} (t) \;\;\; & = & Q^{(+)}_2 (t) \;\;\; + \; Q^{(-)}_2 (t) \quad  = \;   Q^{(+)}_2 (0) \; e^{-\,i\,\Omega_2 t}\; \;\;\,+ \; Q^{(-)}_2 (0) \; e^{+\,i\,\Omega_2 t} \; , \nonumber \\[2mm]
 Q_{1} (t)^+ & = & Q^{(+)}_1 (t)^+ \; + \; Q^{(-)}_1 (t)^+ \; = \;   Q^{(+)}_1 (0)^+ \; e^{+\,i\,\Omega^\ast_1 t}\; + \; Q^{(-)}_1 (0)^+ \; e^{-\,i\,\Omega^\ast_1 t} \; , \nonumber \\[2mm]
 Q_{2} (t)^+ & = & Q^{(+)}_2 (t)^+ \; + \; Q^{(-)}_2 (t)^+ \; = \;   Q^{(+)}_2 (0)^+ \; e^{+\,i\,\Omega^\ast_2 t}\; + \; Q^{(-)}_2 (0)^+ \; e^{-\,i\,\Omega^\ast_2 t}  \; . \label{qsolut2}
 \end{eqnarray}
Due to Eqs.\ (\ref{xmom1}) and (\ref{xmom2}) the canonical conjugate momenta are determined to be:

\begin{eqnarray} P_{1} (t)\;\;\; & = & P^{(+)}_1 (t) \; \;\; + \; P^{(-)}_1 (t) \quad  = \; -\, i\, M_1\Omega_1 \left(  Q^{(+)}_1 (0) \; e^{-\,i\,\Omega_1 t}\;\;\;\, - \; Q^{(-)}_1 (0) \; e^{+\,i\,\Omega_1 t} \right) \; , \nonumber \\[2mm]
 P_{2} (t) \;\;\; & = & P^{(+)}_2 (t) \;\;\; + \; P^{(-)}_2 (t) \quad  = \;  -\, i\, M_2\Omega_2 \left( Q^{(+)}_2 (0) \; e^{-\,i\,\Omega_2 t}\; \;\;\,- \; Q^{(-)}_2 (0) \; e^{+\,i\,\Omega_2 t} \right) \; , \nonumber \\[2mm]
 P_{1} (t)^+ & = & P^{(+)}_1 (t)^+ \; + \; P^{(-)}_1 (t)^+ \; = \;   +\, i\, M^\ast_1\Omega^\ast_1 \left(  Q^{(+)}_1 (0)^+ \; e^{+\,i\,\Omega^\ast_1 t}\; - \; Q^{(-)}_1 (0)^+ \; e^{-\,i\,\Omega^\ast_1 t} \right) \; , \nonumber \\[2mm]
 P_{2} (t)^+ & = & P^{(+)}_2 (t)^+ \; + \; P^{(-)}_2 (t)^+ \; = \;  +\, i\, M^\ast_2\Omega^\ast_2 \left(  Q^{(+)}_2 (0)^+ \; e^{+\,i\,\Omega^\ast_2 t}\; - \; Q^{(-)}_2 (0)^+ \; e^{-\,i\,\Omega^\ast_2 t} \right)\;  .\quad \; \label{psolut2}
 \end{eqnarray}
Canonical quantization requires now the following non-vanishing commutation relations:

\begin{equation} [\,Q_1 \,,\, P_1\, ]\; = \;  i \hbar \; , \quad [\,Q_2 \,,\, P_2\, ]\; = \;  i \hbar \; , \quad   [\,Q^+_1 \,,\, P^+_1\, ]\; = \;  i \hbar \; ,  \quad  [\,Q^+_2 \,,\, P^+_2\, ]\; = \;  i \hbar \; ,\end{equation}
implying

\begin{eqnarray} \frac{2 M_1 \Omega_1}{\hbar}\, [\,Q_1^{(+)}\,,\, Q_1^{(-)}\, ] \quad\;\; & = & 1 \; , \qquad   \frac{2 M_1 \Omega_1}{\hbar}\, [\,Q_2^{(+)}\,,\, Q_2^{(-)}\, ]\quad\;\;\; = \; 1 \; , \nonumber \\[2mm]
  \frac{2 M^\ast_1 \Omega^\ast_1}{\hbar}\, [\,Q_1^{(-)\,+}\,,\, Q_1^{(+)\,+}\, ] & = & 1\; , \qquad  \frac{2 M^\ast_2 \Omega^\ast_2}{\hbar}\, [\,Q_2^{(-)\, +}\,,\, Q_2^{(+)\,+}\, ]\; = \; 1 \; . \label{nonvan1}\end{eqnarray}
In performing the  identifications

\begin{eqnarray} Q^{(+)}_1(0)\;\; &= &  N_1  \; \left(\frac{\hbar}{2\,M_1\Omega_1}\right)^{\frac{1}{2}} \; a_1\; \; , \quad  Q^{(+)}_2(0) \;\;\;\, = \;  N_2  \; \left(\frac{\hbar}{2\,M_2\Omega_2}\right)^{\frac{1}{2}} \;\; a_2 \; , \nonumber\\
 Q^{(-)}_1(0) \;\; &= &  \frac{1}{N_1}  \; \left(\frac{\hbar}{2\,M_1\Omega_1}\right)^{\frac{1}{2}} \; c^+_1 \; , \quad  Q^{(-)}_2(0) \;\;\;\, = \; \frac{1}{N_2}  \; \left(\frac{\hbar}{2\,M_2\Omega_2}\right)^{\frac{1}{2}} \; c^+_2 \; , \nonumber\\
 Q^{(+)}_1(0)^+ &= &  N^\ast_1  \; \left(\frac{\hbar}{2\,M^\ast_1\Omega^\ast_1}\right)^{\frac{1}{2}} \; a^+_1\;  , \quad  Q^{(+)}_2(0)^+ \; = \;  N^\ast_2  \; \left(\frac{\hbar}{2\,M^\ast_2\Omega^\ast_2}\right)^{\frac{1}{2}} \; a^+_2 \; , \nonumber\\ 
 Q^{(-)}_1(0)^+ &= &  \frac{1}{N^\ast_1}  \; \left(\frac{\hbar}{2\,M^\ast_1\Omega^\ast_1}\right)^{\frac{1}{2}} \; c_1 \; , \quad  Q^{(-)}_2(0)^+ \; = \; \frac{1}{N^\ast_2}  \; \left(\frac{\hbar}{2\,M^\ast_2\Omega^\ast_2}\right)^{\frac{1}{2}} \; c_2 \; , \label{identif1}  \end{eqnarray}
the non-vanishing commutation relations Eq.\ (\ref{nonvan1}) reduce to
\begin{eqnarray}  [\,a_1 \,,\, c_1^+\, ]  & = & 1 \; , \qquad    [\,a_2\,,\, c^+_2\, ]\; = \; 1 \; , \nonumber \\[2mm]
 [\,c_1\,,\, a^+_1\, ] & = & 1\; , \qquad   [\,c_2\,,\, a^+_2\, ]\; = \; 1 \; , \label{nonvan2}\end{eqnarray}
coinciding with the expression found in Eqs. (\ref{comrel1}) and (\ref{comrel2}). 

For later convenience the solutions of the Lagrange equations of motion and their canonical conjugate momenta can --- by applying the identifications Eq.\ (\ref{identif1}) to Eqs.\ (\ref{qsolut2}) and (\ref{psolut2}) --- be expressed  in the following form:
\begin{eqnarray} Q_{1} (t)\;\;\; & = & Q^{(+)}_1 (t) \; \;\; + \; Q^{(-)}_1 (t) \quad  = \;   \quad\; \,\left(\frac{\hbar}{2\,M_1\Omega_1}\right)^{\frac{1}{2}} \left( N_1\; a_1 \; e^{-\,i\,\Omega_1 t}\; + \; \frac{c_1^+}{N_1} \; e^{+\,i\,\Omega_1 t}\right) \; , \nonumber \\
 Q_{2} (t) \;\;\; & = & Q^{(+)}_2 (t) \;\;\; + \; Q^{(-)}_2 (t) \quad  = \;  \quad\; \,\left(\frac{\hbar}{2\,M_2\Omega_2}\right)^{\frac{1}{2}} \left( N_2\; a_2 \; e^{-\,i\,\Omega_2 t}\; + \; \frac{c_2^+}{N_2} \; e^{+\,i\,\Omega_2 t}\right)  \; , \nonumber \\
 Q_{1} (t)^+ & = & Q^{(+)}_1 (t)^+ \; + \; Q^{(-)}_1 (t)^+ \; = \;   \quad\; \,\left(\frac{\hbar}{2\,M^\ast_1\Omega^\ast_1}\right)^{\frac{1}{2}}\left(  \frac{c_1}{N^\ast_1} \; e^{-\,i\,\Omega^\ast_1 t}\;\;\;\; + \; N^\ast_1\; a^+_1 \; e^{+\,i\,\Omega^\ast_1 t}\right) \; , \nonumber \\
 Q_{2} (t)^+ & = & Q^{(+)}_2 (t)^+ \; + \; Q^{(-)}_2 (t)^+ \; = \;  \quad\; \,\left(\frac{\hbar}{2\,M^\ast_2\Omega^\ast_2}\right)^{\frac{1}{2}} \left(  \frac{c_2}{N^\ast_2} \; e^{-\,i\,\Omega^\ast_2 t}\;\;\;\; + \; N^\ast_2\; a^+_2 \; e^{+\,i\,\Omega^\ast_2 t}\right)  \; , \nonumber \\
P_{1} (t)\;\;\; & = & P^{(+)}_1 (t) \; \;\; + \; P^{(-)}_1 (t) \quad  = \; -\, i\left(\frac{\hbar\,\Omega_1 M_1}{2}\right)^{\frac{1}{2}} \left( N_1\; a_1 \; e^{-\,i\,\Omega_1 t}\; - \; \frac{c_1^+}{N_1} \; e^{+\,i\,\Omega_1 t}\right) \; , \nonumber \\
 P_{2} (t) \;\;\; & = & P^{(+)}_2 (t) \;\;\; + \; P^{(-)}_2 (t) \quad  = \;  -\, i\left(\frac{\hbar\,\Omega_2 M_2}{2}\right)^{\frac{1}{2}} \left( N_2\; a_2 \; e^{-\,i\,\Omega_2 t}\; - \; \frac{c_2^+}{N_2} \; e^{+\,i\,\Omega_2 t}\right) \; , \nonumber \\
 P_{1} (t)^+ & = & P^{(+)}_1 (t)^+ \; + \; P^{(-)}_1 (t)^+ \; = \;   -\, i\left(\frac{\hbar\,\Omega^\ast_1 M^\ast_1}{2}\right)^{\frac{1}{2}} \left(  \frac{c_1}{N^\ast_1} \; e^{-\,i\,\Omega^\ast_1 t}\;\;\;\; - \; N^\ast_1\; a^+_1 \; e^{+\,i\,\Omega^\ast_1 t}\right) \; , \nonumber \\
 P_{2} (t)^+ & = & P^{(+)}_2 (t)^+ \; + \; P^{(-)}_2 (t)^+ \; = \;  -\, i\left(\frac{\hbar\,\Omega^\ast_2 M^\ast_2}{2}\right)^{\frac{1}{2}} \left(  \frac{c_2}{N^\ast_2} \; e^{-\,i\,\Omega^\ast_2 t}\;\;\;\; - \; N^\ast_2\; a^+_2 \; e^{+\,i\,\Omega^\ast_2 t}\right)\;  .\quad \; \nonumber \\
\label{qpsolut3}
 \end{eqnarray}
The holomorphic, anti-holomorphic and PT-symmtric Hamilton operators $H_H$, $H_A$ and $H$ respectively are obtained from the respective Lagrange functions $L_H$, $L_A$ and $L$ in the standard way by performing a Legendre transformation:

\begin{eqnarray} H_H [Q_1,Q_2\;,P_1,P_2] \;\;\;& = & P_1 \,\; \dot{Q}_1 \; +   P_2 \,\; \dot{Q}_2 \; - \,L_H[Q_1,Q_2\;,\dot{Q}_1,\dot{Q}_2] \; ,  \\[2mm]
H_A [Q^+_1,Q^+_2,P^+_1,P^+_2] & = &\dot{Q}^+_1 \, P^+_1 +  \dot{Q}^+_2\, P_2^+ - \,L_A[Q^+_1,Q^+_2,\dot{Q}^+_1,\dot{Q}^+_2] \; ,  \\[2mm]
H[Q_1,Q_2\;,P_1,P_2,Q^+_1,Q^+_2,P^+_1,P^+_2] & = & H_H [Q_1,Q_2\;,P_1,P_2] + H_A [Q^+_1,Q^+_2,P^+_1,P^+_2] \nonumber \\[2mm]
 & = & P_1 \,\; \dot{Q}_1 \; +   P_2 \,\; \dot{Q}_2 \, + \,\dot{Q}^+_1 \, P^+_1 +  \dot{Q}^+_2\, P_2^+  \nonumber \\[2mm]
 &  &  -\, L_H[Q_1,Q_2\;,\dot{Q}_1,\dot{Q}_2] \, - \,L_A[Q^+_1,Q^+_2,\dot{Q}^+_1,\dot{Q}^+_2]  \nonumber \\[2mm]
 & = & P_1 \,\; \dot{Q}_1 \; +   P_2 \,\; \dot{Q}_2 \, + \,\dot{Q}^+_1 \, P^+_1 +  \dot{Q}^+_2\, P_2^+  \nonumber \\[2mm]
 &  &  -\, L[Q_1,Q_2,\dot{Q}_1,\dot{Q}_2,Q^+_1,Q^+_2,\dot{Q}^+_1,\dot{Q}^+_2]  \; . \; \qquad
\end{eqnarray}
In applying Eqs.\ (\ref{lagr1}), (\ref{lagr2}), (\ref{lagr3}), (\ref{xmom1}) and (\ref{xmom2}) to these identities we obtain:
\begin{eqnarray}  H_H [Q_1,Q_2\;,P_1,P_2] \;\;\; & = & \frac{P_1(t)^2}{2\, M_1}  \;\;\, +  \frac{M_1\,\Omega^2_1}{2} \; Q_1(t)^2 \quad \;\;\, + \; \frac{P_2(t)^2}{2\, M_2} \;\;\, +  \frac{M_2\,\Omega^2_2}{2} \; Q_2(t)^2 \; , \nonumber \\
  H_A [Q^+_1,Q^+_2,P^+_1,P^+_2]  & = & \frac{P_1(t)^{+\,2}}{2\, M^\ast_1}  +  \frac{M^\ast_1\,\Omega^{\ast\, 2}_1}{2} \; Q_1(t)^{+\,2} \; + \; \frac{P_2(t)^{+\,2}}{2\, M^\ast_2}  +  \frac{M^\ast_2\,\Omega^{\ast\,2}_2}{2} \; Q_2(t)^{+\,2} \; ,  \qquad
\end{eqnarray}
and
\begin{eqnarray}
\lefteqn{H[Q_1,Q_2\;,P_1,P_2,Q^+_1,Q^+_2,P^+_1,P^+_2] \; =} \nonumber \\[2mm]
 & = &  \frac{P_1(t)^2}{2\, M_1}  \;\;\, +  \frac{M_1\,\Omega^2_1}{2} \; Q_1(t)^2 \quad \;\;\, + \; \frac{P_2(t)^2}{2\, M_2} \;\;\, +  \frac{M_2\,\Omega^2_2}{2} \; Q_2(t)^2 \nonumber \\
 & + &  \frac{P_1(t)^{+\,2}}{2\, M^\ast_1}  +  \frac{M^\ast_1\,\Omega^{\ast\, 2}_1}{2} \; Q_1(t)^{+\,2} \; + \; \frac{P_2(t)^{+\,2}}{2\, M^\ast_2}  +  \frac{M^\ast_2\,\Omega^{\ast\,2}_2}{2} \; Q_2(t)^{+\,2} \; . \qquad\qquad\qquad\qquad\qquad
\end{eqnarray}
With the help of Eq.\ (\ref{qpsolut3}) the Hamilton operators $H_H$, $H_A$ and $H$ can be expressed in terms of creation and annihilation operators. The result for $H_H$ and $H_A$ --- displaying their time independence --- has been already provided respectively by Eqs.\ (\ref{xyham1}) and (\ref{xyham2}). The --- also time independent ---  PT-symmetric Hamilton operator $H$ being the sum of $H_H$ and $H_A$ is given by
\begin{eqnarray} H \; = \; H_H\,+ \, H_A& = &  \frac{\hbar\, \Omega_1}{2}\left( c_1^+  a_1 + a_1 \,c^+_1\right) \; + \; \frac{\hbar\, \Omega_2}{2}\left( c_2^+ a_2 + a_2 \,c^+_2\right) \nonumber \\
 & + & \frac{\hbar\, \Omega^\ast_1}{2}\left( a_1^+  c_1 + c_1 \,a^+_1\right) \; + \; \frac{\hbar\, \Omega^\ast_2}{2}\left( a_2^+ c_2 + c_2 \,a^+_2\right)  \; .
\end{eqnarray}
In using the commutation relations introduced above and the time-independent Hamilton operator $H=H_H + H_A$  it is simple to show that Eqs.\ (\ref{qsolut2}), (\ref{psolut2}) and (\ref{qpsolut3}) are equivalent to the following identities and underlying Heisenberg equations of motion:
\begin{eqnarray} Q_1(t) \;\;\,  & = & \exp\left(-\frac{Ht}{i\hbar}\right) \; Q_1(0) \; \;\;\exp\left(+\frac{Ht}{i\hbar}\right)\quad  \Leftrightarrow \quad \dot{Q}_1(t) \;\;\, \; = \; -\, \frac{1}{i\hbar} \left[ \; H \, , \, Q_1(t) \; \right]  \; , \nonumber  \\[1mm]
 Q_2(t) \;\;\,  & = & \exp\left(-\frac{Ht}{i\hbar}\right) \; Q_2(0) \; \;\;\exp\left(+\frac{Ht}{i\hbar}\right)\quad  \Leftrightarrow \quad \dot{Q}_2(t) \;\;\, \; = \; -\, \frac{1}{i\hbar} \left[ \; H \, , \, Q_2(t) \; \right]  \; , \nonumber  \\[1mm]
Q_1(t)^+ & = & \exp\left(-\frac{Ht}{i\hbar}\right) \; Q_1(0)^+ \; \exp\left(+\frac{Ht}{i\hbar}\right) \quad \Leftrightarrow \quad \dot{Q}_1(t)^+ \; = \; -\, \frac{1}{i\hbar} \left[ \, H \, , \; Q_1(t)^+ \; \right]\; ,  \nonumber  \\[1mm]
Q_2(t)^+ & = & \exp\left(-\frac{Ht}{i\hbar}\right) \; Q_2(0)^+ \; \exp\left(+\frac{Ht}{i\hbar}\right) \quad \Leftrightarrow \quad \dot{Q}_2(t)^+ \; = \; -\, \frac{1}{i\hbar} \left[ \, H \, , \; Q_2(t)^+ \; \right]\; ,  \nonumber  \\[1mm]
P_1(t) \;\;\, & = & \exp\left(-\frac{Ht}{i\hbar}\right) \; P_1(0) \;\;\; \exp\left(+\frac{Ht}{i\hbar}\right) \quad  \Leftrightarrow  \quad   \dot{P}_1(t) \;\;\,  \;\; = \; -\, \frac{1}{i\hbar} \left[ \; H \, , \, P_1(t) \; \right]\; ,   \nonumber  \\[1mm]
P_2(t) \;\;\, & = & \exp\left(-\frac{Ht}{i\hbar}\right) \; P_2(0) \;\;\; \exp\left(+\frac{Ht}{i\hbar}\right) \quad  \Leftrightarrow  \quad   \dot{P}_2(t) \;\;\,  \; \;= \; -\, \frac{1}{i\hbar} \left[ \; H \, , \, P_2(t) \; \right]\; ,   \nonumber  \\[1mm]
P_1(t)^+ & = & \exp\left(-\frac{Ht}{i\hbar}\right) \; P_1(0)^+ \; \exp\left(+\frac{Ht}{i\hbar}\right)\quad \Leftrightarrow \quad \dot{P}_1(t)^+ \;\; = \; -\, \frac{1}{i\hbar} \left[ \, H \, , \; P_1(t)^+ \; \right]\; ,   \nonumber  \\[1mm]
P_2(t)^+ & = & \exp\left(-\frac{Ht}{i\hbar}\right) \; P_2(0)^+ \; \exp\left(+\frac{Ht}{i\hbar}\right)\quad \Leftrightarrow \quad \dot{P}_2(t)^+ \;\; = \; -\, \frac{1}{i\hbar} \left[ \, H \, , \; P_2(t)^+ \; \right] \; .   
\end{eqnarray}
\section{Path integral quantization of the PT-symmetric Bosonic Harmonic Oscillator}
In order to simplify the notation we want to consider in this section --- without loss of generality --- just one PT-symmetric Bosonic Harmonic oscillator described --- after quantization --- by the Lagrange operators
\begin{eqnarray} L[Q,\dot{Q},Q^+\!,\dot{Q}^+] & = &  L_H[Q,\dot{Q}] + L_A[Q^+,\dot{Q}^+] \nonumber \\[2mm]
 & = & \frac{M}{2} \left( \dot{Q} (t)^2 \, - \, \Omega^2  \, Q(t)^2\right) \; + \; \frac{M^\ast}{2} \left( \dot{Q} (t)^{+\,2} \, - \, \Omega^{\ast\, 2}  \, Q(t)^{+\, 2}\right)  \; ,
\end{eqnarray} 
and  the respective --- time-independent --- Hamilton operators
\begin{eqnarray} H[Q,P,Q^+\!,P^+] & = &  H_H[Q,P] + H_A[Q^+\!,P^+] \nonumber \\[2mm]
 & = & \frac{P^2}{2\, M}  + \frac{M\Omega}{2} \; Q^2 \;\; + \; \frac{P^{+\,2}}{2\, M^\ast}  +  \frac{M^\ast\Omega^\ast}{2} \; Q^{+\,2} \nonumber \\[2mm]
 & = &  \frac{\hbar\, \Omega}{2}\left( c^+  a + a \;c^+\right) \; + \; \frac{\hbar\, \Omega^\ast}{2}\left( a^+ c + c \;a^+\right) \; .
\end{eqnarray} 
The right eigen-states $\big|m,n\big>$ and left eigenstates $\big<\!\big<m,n\big|$ of the Hamilton operator $H$ related to the eigen-values
\begin{equation} E_{mn} \; = \; E_m + E_n^\ast \; = \; \hbar \, \Omega \left( m + \frac{1}{2}\right) \; + \; \hbar \, \Omega^\ast  \left( n + \frac{1}{2}\right)
\end{equation}
with $m$ and $n$ being non-negative integers form a bi-orthonormal and complete basis, i.\ e.,
\begin{equation} \big<\!\big<m,n\big|m^\prime,n^\prime\big> \; = \; \delta_{m,m^\prime} \;  \delta_{n,n^\prime} \; , \qquad I \; = \; \sum\limits_{mn} \big|m,n\big>\big<\!\big<m,n\big| \; . \end{equation}
Hence, the spectral expansion of the Hamilton operator is given by
\begin{equation} H \; = \; \sum\limits_{mn} \big|m,n\big> \, E_{m,n}\,\big<\!\big<m,n\big| \; = \; \sum\limits_{mn} \big|m,n\big> \, E_{m}\,\big<\!\big<m,n\big| + \sum\limits_{mn} \big|m,n\big> \, E^\ast_{n}\,\big<\!\big<m,n\big| \; = \; H_H  +  H_A \; .\end{equation}
In order to proceed in deriving the path integral for this PT-symmetric Bosonic Harmonic oscillator we introduce for convenience the following identities and definitions on the level of Heisenberg operators:

\begin{eqnarray} I  \;\;\, & = &   \exp\left(-\frac{Ht}{i\hbar}\right) \;  I \;\exp\left(+\frac{Ht}{i\hbar}\right) \nonumber \\
 & = &  \int d\zeta\int d\zeta^\ast \; \exp\left(-\frac{Ht}{i\hbar}\right) \;\big|\zeta,\zeta^\ast \big> \; \big<\!\big<\zeta,\zeta^\ast\big| \; \exp\left(+\frac{Ht}{i\hbar}\right) \nonumber \\ 
 & \equiv &  \int d\zeta\int d\zeta^\ast \; \big|\zeta,\zeta^\ast \,  ; \, t \big> \; \big<\!\big<\zeta,\zeta^\ast\, ; \, t  \big|   \; , \\[1mm]
 Q(t) \;\;\, & = &   \exp\left(-\frac{Ht}{i\hbar}\right) \; Q(0) \;\exp\left(+\frac{Ht}{i\hbar}\right) \nonumber \\
 & = &  \int d\zeta\int d\zeta^\ast \; \exp\left(-\frac{Ht}{i\hbar}\right) \;\big|\zeta,\zeta^\ast \big> \; \zeta \; \big<\!\big<\zeta,\zeta^\ast\big| \; \exp\left(+\frac{Ht}{i\hbar}\right) \nonumber \\
 & \equiv &  \int d\zeta\int d\zeta^\ast \; \big|\zeta,\zeta^\ast \,  ; \, t \big> \; \zeta \; \big<\!\big<\zeta,\zeta^\ast\, ; \, t  \big|   \; , \\[1mm]
 Q(t)^+ \;\;\, & = &   \exp\left(-\frac{Ht}{i\hbar}\right) \; Q(0)^+  \;\exp\left(+\frac{Ht}{i\hbar}\right) \nonumber \\
 & = &  \int d\zeta\int d\zeta^\ast \; \exp\left(-\frac{Ht}{i\hbar}\right) \;\big|\zeta,\zeta^\ast \big> \; \zeta^\ast \, \big<\!\big<\zeta,\zeta^\ast\big| \; \exp\left(+\frac{Ht}{i\hbar}\right) \nonumber \\
 & \equiv &  \int d\zeta\int d\zeta^\ast \; \big|\zeta,\zeta^\ast \,  ; \, t \big> \; \zeta^\ast \, \big<\!\big<\zeta,\zeta^\ast\, ; \, t  \big|   \; ,
\end{eqnarray}
and 

\begin{eqnarray}  I \;\;\, & = &   \exp\left(-\frac{Ht}{i\hbar}\right) \;  I \;\exp\left(+\frac{Ht}{i\hbar}\right) \nonumber \\[1mm]
 & = &   \int \frac{dp}{2 \pi  \hbar} \; \int \frac{dp^\ast}{2 \pi  \hbar } \; \exp\left(-\frac{Ht}{i\hbar}\right) \;\big|\,p,p^\ast \big> \; \big<\!\big<p,p^\ast\big| \; \exp\left(+\frac{Ht}{i\hbar}\right) \\[1mm]
 & \equiv &   \int \frac{dp}{2 \pi  \hbar} \; \int \frac{dp^\ast}{2 \pi  \hbar } \; \big|\,p,p^\ast \,  ; \, t \big> \; \big<\!\big<p,p^\ast\, ; \, t  \big|    \; , \nonumber \\[1mm]
P(t) \;\;\, & = &   \exp\left(-\frac{Ht}{i\hbar}\right) \; P(0) \;\exp\left(+\frac{Ht}{i\hbar}\right) \nonumber \\[1mm]
 & = &   \int \frac{dp}{2 \pi  \hbar} \; \int \frac{dp^\ast}{2 \pi  \hbar } \; \exp\left(-\frac{Ht}{i\hbar}\right) \;\big|\,p,p^\ast \big> \; p \; \big<\!\big<p,p^\ast\big| \; \exp\left(+\frac{Ht}{i\hbar}\right) \nonumber \\[1mm]
 & \equiv &   \int \frac{dp}{2 \pi  \hbar} \; \int \frac{dp^\ast}{2 \pi  \hbar } \; \big|\,p,p^\ast \,  ; \, t \big> \; p \; \big<\!\big<p,p^\ast\, ; \, t  \big|   \; , \\[1mm]
 P(t)^+ \;\;\, & = &   \exp\left(-\frac{Ht}{i\hbar}\right) \; P(0)^+  \;\exp\left(+\frac{Ht}{i\hbar}\right) \nonumber \\[1mm]
 & = &   \int \frac{dp}{2 \pi  \hbar} \; \int \frac{dp^\ast}{2 \pi  \hbar } \; \exp\left(-\frac{Ht}{i\hbar}\right) \;\big|\,p,p^\ast \big> \; p^\ast \, \big<\!\big<p,p^\ast\big| \; \exp\left(+\frac{Ht}{i\hbar}\right) \nonumber \\[1mm]
 & \equiv &   \int \frac{dp}{2 \pi  \hbar} \; \int \frac{dp^\ast}{2 \pi  \hbar } \; \big|\,p,p^\ast \,  ; \, t \big> \; p^\ast \, \big<\!\big<p,p^\ast\, ; \, t  \big|   \; .
\end{eqnarray}
With the help of these identities the here so-called {\em generalized retarded Feynman kernel} $\big(\!\big(\zeta^\prime ,\zeta^{\ast\, \prime}\, ; \, t^\prime\,  \big|\,\zeta,\zeta^\ast \,  ; \, t \big)$ can be defined in the standard way \cite{Feynman:1948ur}\cite{Feynman:1949zx}\cite{Dittrich:1992et}\cite{Greiner:1996zu}\cite{Ryder:1985wq}\cite{Kleinert:1993}\cite{Bagarello:2020jhq}, i.~e.,

\begin{equation} \big(\!\big(\zeta^\prime ,\zeta^{\ast\, \prime}\, ; \, t^\prime\,  \big|\,\zeta,\zeta^\ast \,  ; \, t \big) \; \equiv \; \big<\!\big<\zeta^\prime ,\zeta^{\ast\, \prime}\, ; \, t^\prime\,  \big|\,\zeta,\zeta^\ast \,  ; \, t \big> \;\, \theta (t^\prime - t) \; , \end{equation}
with
\begin{eqnarray} \big<\!\big<\zeta^\prime ,\zeta^{\ast\, \prime}\, ; \, t^\prime\,  \big|\,\zeta,\zeta^\ast \,  ; \, t \big>  & = & \big<\!\big<\zeta^\prime ,\zeta^{\ast\, \prime}  \big| \; \exp\left(\frac{H(t^\prime -t)}{i\hbar}\right)\; \big|\zeta,\zeta^\ast \big> \nonumber \\[2mm]
 & = & \sum\limits_{m,n}  \; \big<\!\big<\zeta^\prime ,\zeta^{\ast\, \prime}  \big|m,n \big> \; \exp\left(\frac{E_{m,n}(t^\prime -t)}{i\hbar}\right)\;\big<\!\big<m ,n \big|\zeta,\zeta^\ast \big>\nonumber \\[2mm] 
 & = & \big<\!\big<\zeta^\prime \, , \, t^\prime\,  \big|\,\zeta \,  , \, t \big> \; \big<\zeta^{\ast\, \prime}\, , \, t^\prime \,\big|\,\zeta^\ast \,  , \, t \big>\!\big> \; ,  \qquad \end{eqnarray}
and
\begin{eqnarray} \big<\!\big<\zeta^\prime \, , \, t^\prime\,  \big|\,\zeta \,  , \, t \big>\;\;\,  & \equiv &  \big<\!\big<\zeta^\prime  \big| \; \exp\left(\frac{H_H(t^\prime -t)}{i\hbar}\right)\; \big|\zeta \big> \nonumber \\[2mm]
 & = & \sum\limits_{m}  \; \big<\!\big<\zeta^\prime \big|m \big> \; \exp\left(\frac{E_{m} \,(t^\prime -t)}{i\hbar}\right)\;\big<\!\big<m \big|\zeta \big> \; ,  \\[2mm]
\big<\zeta^{\ast\, \prime}\, , \, t^\prime \,\big|\,\zeta^\ast \,  , \, t \big>\!\big> & \equiv  & \big<\zeta^{\ast\, \prime}  \big| \; \exp\left(\frac{H_A(t^\prime -t)}{i\hbar}\right)\; \big|\zeta^\ast \big>\!\big> \; = \;  \left( \big<\!\big<\zeta \big| \; \exp\left(-\,\frac{H_H(t^\prime -t)}{i\hbar}\right)\; \big|\zeta^\prime  \big>\right)^+ \nonumber \\[2mm]
 & = &   \sum\limits_{n}  \; \big<\zeta^{\ast\, \prime}  \big|n \big>\!\big> \; \exp\left(\frac{E^\ast_{n}\,(t^\prime -t)}{i\hbar}\right)\;\big<n \big|\zeta^\ast \big>\!\big> \; . 
\end{eqnarray}
In considering the definition of the generalized retarded Feynman kernel several remarks are in order: it is clear that the Feynman kernel in the limit $t^\prime -t \rightarrow \pm \,\infty $ should be understood as a distribution. Already at $t^\prime -t  = 0$ there holds due to $\big<\!\big<\zeta^\prime \, , \, t\,  \big|\,\zeta \,  , \, t \big> = \delta (\zeta^\prime - \zeta)$ and $\big<\zeta^{\ast\, \prime}\, , \, t \,\big|\,\zeta^\ast \,  , \, t \big>\!\big> = \delta (\zeta^{\ast\, \prime} - \zeta^\ast)$ the identity
\begin{equation}\big<\!\big<\zeta^\prime ,\zeta^{\ast\, \prime}\, ; \, t\,  \big|\,\zeta,\zeta^\ast \,  ; \, t \big> \; = \; \delta (\zeta^\prime - \zeta) \;  \delta (\zeta^{\ast\, \prime} - \zeta^\ast) \;.
\end{equation}
When choosing causal boundary conditions by claiming Im$[E_m]<0$ we expect $\big<\!\big<\zeta^\prime \, , \, t^\prime\,  \big|\,\zeta \,  , \, t \big>$ to converge and $\big<\zeta^{\ast\, \prime}\, , \, t^\prime \,\big|\,\zeta^\ast \,  , \, t \big>\!\big>$ to diverge for $t^\prime -t  \rightarrow+\, \infty$, while we expect --- inversely --- $\big<\!\big<\zeta^\prime \, , \, t^\prime\,  \big|\,\zeta \,  , \, t \big>$  to diverge and $\big<\zeta^{\ast\, \prime}\, , \, t^\prime \,\big|\,\zeta^\ast \,  , \, t \big>\!\big>$ to converge in the limit $t^\prime -t  \rightarrow -\infty$. In how far the kernel $\big<\!\big<\zeta^\prime ,\zeta^{\ast\, \prime}\, ; \, t^\prime\,  \big|\,\zeta,\zeta^\ast \,  ; \, t \big> =  \big<\!\big<\zeta^\prime \, , \, t^\prime\,  \big|\,\zeta \,  , \, t \big> \; \big<\zeta^{\ast\, \prime}\, , \, t^\prime \,\big|\,\zeta^\ast \,  , \, t \big>\!\big>$ converges or diverges for causal boundary conditions in the limit $t^\prime -t \rightarrow \pm \, \infty$ is yet unclear. Nonetheless it is clear that the generalized retarded Feynman kernel $\big(\!\big(\zeta^\prime ,\zeta^{\ast\, \prime}\, ; \, t^\prime\,  \big|\,\zeta,\zeta^\ast \,  ; \, t \big)$ and the resulting retarded path integral is well defined for the finite time interval $t^\prime - t$, as long as there exist respectively the right and left vacuum states $\big|0\big>$ and $\big<\!\big< 0\big|$ with $\big<\!\big< 0\big|0\big>=1$. 

The derivation of the path integral is normaly achieved \cite{Dittrich:1992et}\cite{Greiner:1996zu}\cite{Ryder:1985wq}\cite{Kleinert:1993}\cite{Bagarello:2020jhq} by choosing an infinity of ordered time values $t = t_0 < t_1 < t_2 < \ldots < t_{N-1} < t_N = t^\prime$ separated by an infinitesimal time interval $\Delta t = (t^\prime - t)/N$ with $N\rightarrow \infty$. The generalized retarded Feynman kernel is then factorised as a product of an infinity of retarded Feynman kernels describing infinitesimal time intervals:
\begin{eqnarray}  \lefteqn{\big(\!\big(\zeta^\prime ,\zeta^{\ast\, \prime}\, ; \, t^\prime\,  \big|\zeta,\zeta^\ast \,  ; \, t \big)  \; = \; \lim\limits_{N\rightarrow \infty}  \int d\zeta_{N-1}\int d\zeta_{N-1}^\ast \; \ldots \;  \int d\zeta_2\int d\zeta_2^\ast \int d\zeta_1\int d\zeta_1^\ast} \nonumber \\[3mm]
 & & \big(\!\big(\zeta^\prime ,\zeta^{\ast\, \prime}\, ; \, t^\prime\,  \big|\zeta_{N-1},\zeta^\ast_{N-1} \,  ; \, t_{N-1} \big) \; \ldots \; \big(\!\big(\zeta_2 ,\zeta^\ast_2\, ; \, t_2\,  \big|\zeta_1,\zeta^\ast_1 \,  ; \, t_1 \big)\big(\!\big(\zeta_1 ,\zeta^\ast_1\, ; \, t_1\,  \big|\zeta,\zeta^\ast \,  ; \, t \big)   \; . \end{eqnarray}
The retarded Feynman kernels $\big(\!\big(\zeta_{k+1} ,\zeta^\ast_{k+1}\, ; \, t_{k+1}\,  \big|\zeta_k,\zeta^\ast_k \,  ; \, t_k \big)  =\big<\!\big<\zeta_{k+1} ,\zeta^\ast_{k+1}\, ; \, t_{k+1}\,  \big|\zeta_k,\zeta^\ast_k \,  ; \, t_k \big> \; \theta ( \Delta t)$ describing an infinitesimal time-interval  $\Delta t = t_{k+1} - t_k$ are typically Taylor expanded in $\Delta t\,$:
\begin{eqnarray}\lefteqn{\big<\!\big<\zeta_{k+1} ,\zeta^\ast_{k+1}\, ; \, t_{k+1}\,  \big|\zeta_k,\zeta^\ast_k \,  ; \, t_k \big> \; = \; \big<\!\big<\zeta_{k+1} ,\zeta^\ast_{k+1}\,  \big| \;\exp\left(\frac{H \Delta t}{i\hbar}\right) \; \big|\zeta_k,\zeta^\ast_k \,  \big> \; =} \nonumber \\[1mm]
 & = & \big<\!\big<\zeta_{k+1} ,\zeta^\ast_{k+1}\,  \big| \; \left( 1 +\,\frac{H \Delta t}{i\hbar}\right) \; \big|\zeta_k,\zeta^\ast_k \,  \big> \; + \; O\Big((\Delta t)^{\,2}\Big)\nonumber \\[1mm]
 & = &  \int \frac{dp_k}{2 \pi  \hbar} \; \int \frac{dp^\ast_k}{2 \pi  \hbar } \;\big<\!\big<\zeta_{k+1} ,\zeta^\ast_{k+1}\,  \big|\,p_k,p^\ast_k\big> \big<\!\big<p_k,p^\ast_k\big|    \left( 1 +\,\frac{H \Delta t}{i\hbar}\right) \; \big|\zeta_k,\zeta^\ast_k \,  \big> \; + \; O\Big((\Delta t)^{\,2}\Big) \nonumber \\[1mm]
 & = &  \int \frac{dp_k}{2 \pi  \hbar} \;  \;\; \exp \left( - \frac{p_k (\zeta_{k+1} - \zeta_k) }{i\hbar} \right)    \left( 1 +\,\frac{H_H(p_k,\zeta_k)}{i\hbar} \; \Delta t\right) \nonumber \\[1mm] 
 &  & \int \frac{dp^\ast_k}{2 \pi  \hbar } \; \;\; \exp \left( - \frac{p^\ast_k (\zeta^\ast_{k+1} - \zeta^\ast_k) }{i\hbar} \right)    \left( 1 +\,\frac{H_A(p^\ast_k,\zeta^\ast_k)}{i\hbar} \; \Delta t\right)+ \; O\Big((\Delta t)^{\,2}\Big)  \; . \quad
\end{eqnarray}
With the help of this expression it is straight forward to reformulate the retarded Feynman kernel in terms of a  here so-called {\em generalized retarded canonical path integral} \cite{Greiner:1996zu}\cite{Ryder:1985wq}\cite{Kleinert:1993}:
\begin{eqnarray} \lefteqn{\big(\!\big(\zeta^\prime ,\zeta^{\ast\, \prime}\, ; \, t^\prime\,  \big|\zeta,\zeta^\ast \,  ; \, t \big) \; =} \nonumber \\[2mm]
 & = & \lim\limits_{N\rightarrow \infty}  \int \, \prod\limits_{k=1}^{N-1} d\zeta_k  \; \prod\limits_{k=0}^{N-1} \frac{dp_k}{2 \pi  \hbar}  \; \exp \left[ \frac{i}{\hbar} \; \Delta t \sum\limits_{k=0}^{N-1} \left( p_k \; \frac{\zeta_{k+1} - \zeta_k}{\Delta t}\, - \, H_H (p_k,\zeta_k)\right)  \right]\nonumber \\[2mm] 
 & & \qquad \int \,  \prod\limits_{k=1}^{N-1} d\zeta^\ast_k \; \prod\limits_{k=0}^{N-1} \frac{dp^\ast_k}{2 \pi  \hbar} \; \exp \left[ \frac{i}{\hbar} \; \Delta t \sum\limits_{k=0}^{N-1} \left( p^\ast_k \; \frac{\zeta^\ast_{k+1} - \zeta^\ast_k}{\Delta t}\, - \, H_A (p^\ast_k,\zeta^\ast_k)\right)  \right] \nonumber \\[2mm]
 & = & \int {\cal D}\zeta \int \frac{{\cal D}p}{h} \;\;  \exp \left[  \frac{i}{\hbar}  \int_t^{t^\prime}\!\! d\tau \; \left( p \; \dot{\zeta} \; \;\;\; - \; H_H (p,\zeta) \right) \; \right] \nonumber \\
 &  &  \int {\cal D} \zeta^\ast\! \int \frac{{\cal D}p^\ast}{h}   \exp \left[  \frac{i}{\hbar}  \int_t^{t^\prime}\!\! d\tau \; \left(  \dot{\zeta}^\ast \,p^\ast \, -\;  H_A (p^\ast,\zeta^\ast) \right) \; \right] \nonumber \\
 & = & \int {\cal D}\zeta \int \frac{{\cal D}p}{h} \int {\cal D} \zeta^\ast\! \int \frac{{\cal D}p^\ast}{h} \; \exp \left[  \frac{i}{\hbar}  \int_t^{t^\prime}\!\! d\tau \; \left( p \; \dot{\zeta} \; + \;  \dot{\zeta}^\ast \, p ^\ast \; - \; H_H (p,\zeta) -\;  H_A (p^\ast,\zeta^\ast) \right) \; \right] .\nonumber \\
\end{eqnarray} 
With the following ansatz for the holomorphic Hamilton function $H_H (p,\zeta)$ and anti-holomorphic Hamilton function $H_A (p^\ast,\zeta^\ast)= \big(H_H (p,\zeta) \big)^\ast$, i.~e.,
\begin{equation} H_H (p,\zeta) \; = \; \frac{p^2}{2\,M} + V_H(\zeta) \; , \quad   H_A (p^\ast,\zeta^\ast) \; = \; \frac{p^{\ast \, 2}}{2\,M^\ast} + V_A(\zeta^\ast) \; = \; \frac{p^{\ast \, 2}}{2\,M^\ast} + \Big( V_H(\zeta) \Big)^\ast \; , \end{equation}
it is possible to perform the Gaussian momentum integration in the retarded canonical path integral leading to the retarded path integral in Feynman's form, i.~e.,
\begin{eqnarray} \lefteqn{\big(\!\big(\zeta^\prime ,\zeta^{\ast\, \prime}\, ; \, t^\prime\,  \big|\zeta,\zeta^\ast \,  ; \, t \big) \; = } \nonumber \\ [2mm]
 & = & {\cal N}\int {\cal D}\zeta \;\;  \exp \left[  \frac{i}{\hbar}  \int_t^{t^\prime}\!\! d\tau \; L_H (\dot{\zeta},\zeta)  \; \right] \;{\cal N}^\ast  \int {\cal D} \zeta^\ast \;   \exp \left[  \frac{i}{\hbar}  \int_t^{t^\prime}\!\! d\tau \;  L_A (\dot{\zeta}^\ast,\zeta^\ast)  \; \right] \nonumber \\[1mm]
 & = &  \left|\, {\cal N}\, \right|^2 \int {\cal D}\zeta \int {\cal D} \zeta^\ast \;\;  \exp \left[  \frac{i}{\hbar}  \int_t^{t^\prime}\!\! d\tau \; \left(  L_H (\dot{\zeta},\zeta) +  L_A (\dot{\zeta}^\ast,\zeta^\ast) \right) \; \right] ,
\end{eqnarray} 
with ${\cal N}$ being some non-vanishing constant and $L_H (\dot{\zeta},\zeta)$ and $L_A (\dot{\zeta}^\ast,\zeta^\ast) = \big( L_H (\dot{\zeta},\zeta)\big)^\ast$ being respectively the holomorphic   and anti-holomorphic Lagrange function given by
\begin{equation} L_H (\dot{\zeta},\zeta) \; = \;  \; \frac{M}{2} \; \dot{\zeta}^{\,2} - V_H(\zeta) \; , \quad   L_A (\dot{\zeta}^\ast,\zeta^\ast) \; = \; \frac{M^\ast}{2} \; \dot{\zeta}^{\ast\, 2} - V_A(\zeta^\ast) \; = \; \frac{M^\ast}{2} \; \dot{\zeta}^{\ast\, 2} - \Big( V_H(\zeta) \Big)^\ast . \end{equation}
The quantity allowing to calculate Feynman Green's \cite{Economou:1979} or vertex functions being quantum expectation values of time-ordered products of Heisenberg operators is known to be \cite{Greiner:1996zu}\cite{Ryder:1985wq} the here so-called {\em generalized retarded vacuum persistence amplitude} $Z[J,J^\ast]= Z_H[J] \; Z_A[J^\ast]$, i.~e.,
\begin{eqnarray}  Z[J\,,J^\ast]  & = & \int {\cal D}\zeta \int \frac{{\cal D}p}{h} \int {\cal D} \zeta^\ast\! \int \frac{{\cal D}p^\ast}{h} \nonumber \\[1mm]
 & & \exp \left[  \frac{i}{\hbar} \int_{-\infty}^{+\infty}\!\! dt \; \left( p \; \dot{\zeta} \; + \;   \dot{\zeta}^\ast \, p^\ast\;  -  H_H (p,\zeta) -  H_A (p^\ast,\zeta^\ast) + J\, \zeta  + \zeta^\ast J^\ast \right) \; \right] \nonumber \\[1mm]
 & = &  \left| \,{\cal N} \, \right|^2 \int {\cal D}\zeta \int {\cal D} \zeta^\ast \;  \exp \left[  \frac{i}{\hbar}  \int_{-\infty}^{+\infty}\!\! dt \; \left(  L_H (\dot{\zeta},\zeta) +  L_A (\dot{\zeta}^\ast,\zeta^\ast) + J\, \zeta  + \zeta^\ast  J^\ast \right) \; \right] , \qquad \\[2mm]
Z_H[J]\;\,  & = &  \int {\cal D}\zeta \int \frac{{\cal D}p}{h}  \;\; \, \exp \left[  \frac{i}{\hbar}  \int_{-\infty}^{+\infty}\!\! dt \; \left( p \; \dot{\zeta} \,   - \, H_H (p,\zeta) + J\, \zeta \right)  \; \right] \nonumber \\[2mm]
 & = &   {\cal N} \; \int {\cal D}\zeta \; \;\,  \exp \left[  \frac{i}{\hbar}  \int_{-\infty}^{+\infty}\!\! dt \; \left(  L_H (\dot{\zeta},\zeta)  + J\, \zeta  \right) \; \right] \, , \\[1mm]
Z_A[J^\ast] & = &  \int {\cal D} \zeta^\ast\! \int \frac{{\cal D}p^\ast}{h}   \exp \left[  \frac{i}{\hbar}  \int_{-\infty}^{+\infty}\!\! dt  \; \left(   \dot{\zeta}^\ast\, p^\ast\, -\,  H_A (p^\ast,\zeta^\ast)  +  \zeta^\ast J^\ast \right) \; \right]  \nonumber  \\[1mm]
 & = &   {\cal N}^\ast\int {\cal D} \zeta^\ast \;  \exp \left[  \frac{i}{\hbar}  \int_{-\infty}^{+\infty}\!\! dt \; \left(   L_A (\dot{\zeta}^\ast,\zeta^\ast)  + \zeta^\ast J^\ast  \right) \; \right] \, . 
\end{eqnarray}
As pointed out before in the context of the generalized retarded Feynman kernel the time-integral $\int_{-\infty}^{+\infty} dt$ should be understood --- with due care --- as the limit of the integral $\int_t^{t^\prime} d\tau$ with $t^\prime - t \rightarrow + \infty$ in order to preserve a well-defined mathematical framework \cite{Brush:1961}. Nonetheless it is clear that the generalized retarded vacuum persistence amplitude should be considered in the same way as the generalized retarded Feynman kernel to be a distribution whose divergence or convergence in the limit of $t^\prime - t \rightarrow  + \infty$ remains yet unclear. We recall at this point again that the integration contours for the variables $\zeta$, $\zeta^\ast$, $p$ and $p^\ast$ should be chosen with due care in order to maintain the eigen-functions of the $PT$-symmetric Hamilton operator $H=H_H+H_A$ square-integrable. In order to guarantee the convergence of the Minkowskian path integral presented here we have to request additionally that the eigen-values $E_m$ of the holomorphic Hamilton operator $H_H=H_A^+$ should possess at least some infinitesimal negative imaginary part, i.\ e.\, Im$[E_m]<0$,  in order to enforce a retarded causal evolution of the system. 

Keeping this discussion in mind it gets clear how to consistently set up ---   for the PT-symmetric Bosonic Harmonic Oscillator  --- the retarded vacuum persistence amplitude $Z[J\,,J^\ast]$ and the so-called {\em generalized retarded generating functional} \cite{Greiner:1996zu}\cite{Ryder:1985wq} for irreducible (connected) Feynman Green's functions $W[J\,,J^\ast] = W_H[J] + W_A[J^\ast]$ being a sum of the so-called  {\em retarded holomorphic generating functional} $W_H[J]$ and {\em retarded anti-holomorphic generating functional} $W_A[J^\ast]$, i.~e.,

\begin{eqnarray} Z[J\,,J^\ast]  & = & \exp\left( \frac{i}{\hbar} \, W[J\,,J^\ast]  \right)  \; = \; \exp\left( \frac{i}{\hbar} \, W_H[J]  \right) \;  \exp\left( \frac{i}{\hbar} \, W_A[J^\ast]  \right)  \nonumber \\[2mm] 
 & = & {\cal N}\,\;  \int {\cal D}\zeta   \; \, \exp \left[  \frac{i}{\hbar}  \int_{-\infty}^{+\infty}\!\! dt \; \left( \frac{M}{2} \; \dot{\zeta}^2\;\; \; \, - \frac{M\,\Omega^2}{2} \;\zeta^2 \;\;\;\;\,\, + \; J\; \zeta \right)  \; \right] \nonumber \\[2mm]
 &  & {\cal N}^\ast  \int {\cal D} \zeta^\ast   \exp \left[  \frac{i}{\hbar}  \int_{-\infty}^{+\infty}\!\! dt  \; \left(  \frac{M^\ast}{2} \; \dot{\zeta}^{\ast\, 2}\, -  \frac{M^\ast\Omega^{\ast\, 2}}{2} \;\zeta^{\ast\,2}    + \; \zeta^\ast J^\ast \right) \; \right] \; .
\end{eqnarray}
The general difference between Classical Theory and Quantum Theory is that the evolution of a quantum system is described by correlators containing more than one Heisenberg operator. The most prominent correlators solving Schwinger-Dyson or renormalization-group equations are retarded vacuum expectation values of time-ordered Heisenberg operators which have been introduced by R.\ P.\ Feynman and which are calculated in the presented formalism by taking variational derivatives of the generalized retarded vacuum persistence amplitude $Z[ J\, ,  J^\ast ]$ \cite{Greiner:1996zu}\cite{Ryder:1985wq}, i.~e.,
\begin{eqnarray} \lefteqn{ \left. \left(\frac{\hbar}{i}\right)^{k+\ell} \frac{\delta^{k+\ell}\, Z[ J\, ,  J^\ast ]}{\delta J^\ast (t_{k+\ell}) \ldots \delta J^\ast(t_{k+1})\, \delta J(t_k) \ldots \delta J(t_1)} \right|_{J=J^\ast = 0} \; =}\nonumber \\[2mm]
 & = & \big<\!\big< 0 \big| \, T [\, Q(t_1) \ldots Q(t_k) \, ] \,\big| 0 \big> \; \Big(\big<\!\big< 0 \big| \, T [\, Q(t_{k+1}) \ldots Q (t_{k+\ell})\, ] \,\big| 0 \big>\Big)^+  \nonumber \\[2mm]
  & = & \left. \left(\frac{\hbar}{i}\right)^{k} \frac{\delta^{k}\, Z_H[ J]}{\delta J(t_k) \ldots \delta J(t_1)} \right|_{J=0} \left. \left(\frac{\hbar}{i}\right)^{\ell} \frac{\delta^{\ell}\, Z_A[  J^\ast ]}{\delta J^\ast (t_{k+\ell}) \ldots \delta J^\ast(t_{k+1})} \right|_{J^\ast = 0} \nonumber \\[2mm]
 & = & G^{(k)}_H (t_1, \ldots, t_k)\; G^{(\ell)}_A (t_{k+1}, \ldots , t_{k+\ell})  \; . \end{eqnarray}
\section{Elements of a  PT-symmetric Quantum Field Theory for Bosons and Fermions}
The generalization of the whole formalism to (non-relativistic and relativistic) non-Hermitian Quantum Field Theory appears straight forward \cite{Greiner:1996zu}\cite{Ryder:1985wq}. For the free neutral Bosonic Nakanishi-Klein-Gordon (NKG) field \cite{Nakanishi:1972wx}\cite{Nakanishi:1972pt}\cite{Kleefeld:2001xd}\cite{Kleefeld:2003zj} described by the PT-symmetric action
\begin{eqnarray} S_{\tiny \mbox{NKG,0}}[\,\phi\,,\phi^\ast\,] & = &   \int d^4x \; \Big[\;\;  \frac{1}{2} \; \Big( \partial_\mu \phi (x) \Big)\;\Big( \partial^\mu \phi (x) \Big)\;\; \, - \, \frac{M^2}{2} \;\phi (x)^2 \nonumber \\
 & & \qquad\;\;\;+ \,\frac{1}{2} \;  \Big( \partial_\mu \phi (x)^\ast \Big)\Big( \partial^\mu \phi (x)^\ast \Big) \, -  \, \frac{M^{\ast\, 2}}{2} \;\phi (x)^{\ast\,2}  \; \Big] 
\end{eqnarray}
the retarded vacuum persistence amplitude $Z_{\tiny \mbox{NKG,0}}[J\,,J^\ast]$ and generating functional $W_{\tiny \mbox{NKG,0}}[J\,,J^\ast]$ take in natural units ($\hbar = c =1$) e.~g.\ the following form:
\begin{eqnarray} \lefteqn{Z_{\tiny \mbox{NKG,0}}[J\,,J^\ast] \; = \; \exp\Big( i \, W_{\tiny \mbox{NKG,0}}[J\,,J^\ast]  \Big)  \; =} \nonumber \\[2mm] 
 & = & \; |\,{\cal N}_{\tiny \mbox{NKG}}|^2 \int {\cal D}\phi  \int {\cal D} \phi^\ast   \; \, \exp \left[ \, i  \,S_{\tiny \mbox{NKG,0}}[\,\phi\,,\phi^\ast\,] + \, i  \int d^4x \; \Big( J(x)\; \phi(x) + \phi^\ast (x) \, J^\ast(x)\Big) \; \right] \nonumber \\[2mm]
 & = & {\cal N}_{\tiny \mbox{NKG}} \int {\cal D}\phi   \; \, \exp \left[ \, i  \int d^4x \; \left( \frac{1}{2} \; \Big( \partial_\mu \phi (x) \Big)\Big( \partial^\mu \phi (x) \Big) \; \;\; \, - \, \frac{M^2}{2} \;\;\,\phi (x)^2 \;\;\;  + \; J(x)\; \phi(x) \right)  \; \right] \nonumber \\[2mm]
 &  & {\cal N}^\ast_{\tiny \mbox{NKG}}  \int {\cal D} \phi^\ast   \exp \left[ \, i  \int d^4x \; \left(  \frac{1}{2} \;  \Big( \partial_\mu \phi (x)^\ast \Big)\Big( \partial^\mu \phi (x)^\ast \Big) \, -  \, \frac{M^{\ast\, 2}}{2} \;\phi (x)^{\ast\,2} \;   + \; \phi^\ast (x) \, J^\ast(x) \right) \; \right] \; , \nonumber \\
\end{eqnarray}
implying in natural units
\begin{eqnarray}  W_{\tiny \mbox{NKG,0}}[J\,,J^\ast] & = & \frac{1}{i} \; \ln Z_{\tiny \mbox{NKG,0}}[J\,,J^\ast] \nonumber \\[2mm]
 & = &  \frac{i^2}{2} \int d^4x_1 \int d^4 x_2 \;  J(x_1)  \;\,  \int \frac{d^4p}{(2\pi)^4} \; e^{-\, i\, p\cdot (x_1 - x_2)}\; \frac{1}{p^2- M^2}\;\; \;  J(x_2) \nonumber \\[2mm]
 & +  & \frac{i^2}{2} \int d^4x_2 \int d^4 x_1 \; J(x_2)^\ast \; \int \frac{d^4p}{(2\pi)^4} \; e^{+\, i\, p\cdot (x_1 - x_2)}\; \frac{1}{p^2- M^{\ast\,2}}\;  J(x_1)^\ast  . \quad
\end{eqnarray}
The corresponding causal and anti-causal retarded Feynman propagators \cite{Kleefeld:2003zj}\cite{Kleefeld:2001xd} are with Im[$\,M\,$]$\,<0$ in natural units ($\,T[\ldots]$ denotes here time-ordering and $\overline{T}[\ldots]$ anti-time-ordering!~):

\begin{eqnarray}  \lefteqn{(-\, i) \, \big<\!\big< 0 \big| \, T [\, \phi(x_1) \, \phi(x_2) \, ] \,\big| 0 \big> \quad  \; = \;  \int \frac{d^4p}{(2\pi)^4} \; e^{-\, i\, p\cdot (x_1 - x_2)}\; \frac{1}{p^2- M^2} \; \,\;=}\nonumber \\[1mm] 
 & = & (-\,i) \; \left( \frac{1}{i} \right)^2 \left. \; \frac{\delta^{2}\,  Z_{\tiny \mbox{NKG,0}}[J\,,J^\ast]}{\delta J(x_2) \, \delta J(x_1) } \right|_{J=J^\ast = 0} \; = \; \left( \frac{1}{i} \right)^2\; \frac{\delta^{2}\,  W_{\tiny \mbox{NKG,0}}[J\,,J^\ast]}{\delta J(x_2) \; \delta J(x_1) } \; ,    \\[3mm] 
\lefteqn{(+\, i) \; \big<\!\big< 0 \big| \, \overline{T} [\, \phi(x_2)^+  \phi(x_1)^+  ] \,\big| 0 \big>  \; = \;  \int \frac{d^4p}{(2\pi)^4} \; e^{+\, i\, p\cdot (x_1 - x_2)}\; \frac{1}{p^2- M^{\ast \, 2}} \; =} \nonumber \\[1mm]
 & = & (+\,i) \; \left( \frac{1}{i} \right)^2 \left. \; \frac{\delta^{2}\,  Z_{\tiny \mbox{NKG,0}}[J\,,J^\ast]}{\delta J(x_1)^\ast \, \delta J(x_2)^\ast } \right|_{J=J^\ast = 0} \; = \; \left( \frac{1}{i} \right)^2 \; \frac{\delta^{2}\,  W_{\tiny \mbox{NKG,0}}[J\,,J^\ast]}{\delta J(x_1)^\ast \, \delta J(x_2)^\ast} \nonumber \\[2mm]
 & = &  (+\, i) \, \Big(\big<\!\big< 0 \big| \, T [\, \phi(x_1) \; \phi(x_2)  ] \,\big| 0 \big>\Big)^+ \; .   
\end{eqnarray}
Analogously the retarded vacuum persistence amplitude $Z_{\tiny \mbox{LWD,0}}[\,\eta\,,\eta^\ast\,]$ and generating functional $W_{\tiny \mbox{LWD,0}}[\,\eta\,,\eta^\ast\,]$ for the free neutral Fermionic Lee-Wick-Dirac (LWD) field \cite{Kleefeld:2003zj}\cite{Kleefeld:1999}\cite{Kleefeld:1998dg}\cite{Kleefeld:1998yj}\cite{Kleefeld:2021sqz}\cite{Lee:1970iw}\cite{Cotaescu:1983nc}
being described by the PT-symmetric action

\begin{eqnarray} S_{\tiny \mbox{LWD,0}}[\,\psi\,,\psi^\ast\,] & = & \int d^4x \; \Bigg[  \frac{1}{2} \;\, \overline{\psi^c(x)}\; \left(  \frac{i}{2}\! \stackrel{\,\leftrightarrow}{\not\!\partial}  - \, M    \right) \;  \psi(x)   \; + \;  \frac{1}{2} \; \, \overline{\psi(x)} \; \; \left(  \frac{i}{2}\! \stackrel{\,\leftrightarrow}{\not\!\partial}  - \, M^\ast    \right) \psi^c(x) \Bigg]  \nonumber \\[2mm]
\end{eqnarray}
reads  in natural units:

\begin{eqnarray} \lefteqn{Z_{\tiny \mbox{LWD,0}}[\,\eta\,,\eta^\ast\,] \; = \; \exp\Big( i \, W_{\tiny \mbox{LWD,0}}[\,\eta\,,\eta^\ast\,]  \Big)  \; =} \nonumber \\[2mm] 
 & = & \; |\,{\cal N}_{\tiny \mbox{LWD}}|^2 \int {\cal D}\psi  \int {\cal D} \psi^\ast   \; \, \exp \left[ \, i  \,S_{\tiny \mbox{LWD,0}}[\,\psi\,,\psi^\ast\,] + \, i  \int d^4x \; \left( \, \overline{\eta^c(x)} \, \psi(x) + \overline{\psi(x)} \, \eta^c(x)\right) \; \right] \nonumber \\[2mm]
 & = & {\cal N}_{\tiny \mbox{LWD}} \int {\cal D}\psi   \; \, \exp \left[ \, i  \int d^4x \; \left( \frac{1}{2} \;\, \overline{\psi^c(x)}\; \left(  \frac{i}{2}\! \stackrel{\,\leftrightarrow}{\not\!\partial}  - \, M    \right) \;  \psi(x) \;  \; + \; \overline{\eta^c(x)} \, \psi(x) \right)  \; \right] \nonumber \\[2mm]
 &  & {\cal N}^\ast_{\tiny \mbox{LWD}}  \int {\cal D} \psi^\ast   \exp \left[ \, i  \int d^4x \; \left(  \frac{1}{2} \; \, \overline{\psi(x)} \; \; \left(  \frac{i}{2}\! \stackrel{\,\leftrightarrow}{\not\!\partial}  - \, M^\ast    \right) \psi^c(x) \;   + \; \overline{\psi(x)}\; \eta^c(x) \right) \; \right] \; .
\end{eqnarray}
Here we used the standard abbreviations $\psi^c(x) = C \, \gamma^0 \, \psi(x)^\ast $, $\overline{\psi^c(x)} = \psi(x)^T  \,C$ and $\overline{\psi(x)} = \psi (x)^+\, \gamma^0$ with $C = - \,C^T = - \, C^+ = - \,C^{-1}  =  i \, \gamma^2\,  \gamma^0$ being the charge conjugation matrix respecting $\gamma^{\mu\, T} = - \,C \, \gamma^\mu \, C^{-1}$. It should be noted that due to the Grassmann nature of the fields and the external sources there holds $\overline{\eta^c(x)} \, \psi(x) = \overline{\psi^c(x)} \, \eta(x)$ and $\overline{\psi(x)}\; \eta^c(x) = \overline{\eta(x)}\; \psi^c(x) $. Keeping in mind that the short-hand notation $\psi (x)$ and $\psi(x)^\ast$ abbreviates that these fields are Fermionic spinor fields possessing four components (implying $\int {\cal D}\psi  \int {\cal D} \psi^\ast =\int {\cal D}\psi^\ast  \int {\cal D} \psi$) it is straight forward to obtain the following result in natural units:

\begin{eqnarray}  \lefteqn{W_{\tiny \mbox{LWD,0}}[\,\eta\,,\eta^\ast \,] \; = \; \frac{1}{i} \; \ln Z_{\tiny \mbox{LWD,0}}[\,\eta\,,\eta^\ast\,] \; = } \nonumber \\[2mm]
 & = &  \frac{i^2}{2} \,  \int d^4x_1 \int d^4 x_2 \;\;  \overline{\eta^c(x_1)}    \int \frac{d^4p}{(2\pi)^4} \; e^{-\, i\, p\cdot (x_1 - x_2)}\;\;\; \frac{\not\! p + M}{p^2- M^2}\;\; \, \eta(x_2) \nonumber \\[2mm]
 & + & \frac{i^2}{2}\,  \int d^4x_2 \int d^4 x_1 \; \; \overline{\eta(x_2)}  \;  \int \frac{d^4p}{(2\pi)^4} \; e^{+\, i\, p\cdot (x_1 - x_2)}\; \; \; \frac{\not\! p + M^\ast}{p^2- M^{\ast\,2}}\;\; \eta^c (x_1) \nonumber \\[2mm]
 & = &  \frac{i^2}{2} \,  \int d^4x_1 \int d^4 x_2 \;\; \sum\limits_{rs} \int \frac{d^4p}{(2\pi)^4} \; e^{-\, i\, p\cdot (x_1 - x_2)}\; \frac{ \eta(x_1)_r \;  [\,  C\, (\not\! p + M)\, ]_{rs}\;\eta(x_2)_s}{p^2- M^2}   \nonumber \\[2mm]
 & + & \frac{i^2}{2}\,  \int d^4x_2 \int d^4 x_1 \; \; \sum\limits_{sr} \int \frac{d^4p}{(2\pi)^4} \; e^{+\, i\, p\cdot (x_1 - x_2)}\; \frac{\eta(x_2)^\ast_s\; [\, \gamma^0 \, (\not\! p + M^\ast)\, C\, \gamma^0\, ]_{sr}\; \eta (x_1)^\ast_r}{p^2- M^{\ast\,2}}\;   . \qquad
\end{eqnarray}
The corresponding (anti-)causal Feynman propagators \cite{Kleefeld:2003zj} in natural units are with Im[$\,M\,$]$\,<0$:
\begin{eqnarray}  \lefteqn{(-\, i) \, \big<\!\big< 0 \big| \, T [\, \Big[\, [C\, \psi(x_1)]_r \, [- \, C\, \psi(x_2)]_s \, \Big] \,\big| 0 \big> \quad  \;\;\, = \; (-\, i) \, \big<\!\big< 0 \big| \, T [\, \Big[\, [C\, \psi(x_1)]_r \, [\,\overline{\psi^c(x_2)}\,]_s \, \Big] \,\big| 0 \big> \; =}\nonumber \\[2mm] 
 & = & (-\,i) \; \left( \frac{1}{i} \right)^2 \left. \; \frac{\delta^{2}\,  Z_{\tiny \mbox{LWD,0}}[\,\eta\,,\eta^\ast]}{\delta \eta(x_2)_s \; \delta \eta (x_1)_r } \right|_{\eta \, =\, \eta^\ast = 0} \; = \; \left( \frac{1}{i} \right)^2\; \frac{\delta^{2}\,  W_{\tiny \mbox{LWD,0}}[\,\eta\,,\eta^\ast]}{\delta \eta(x_2)_s \; \delta \eta (x_1)_r}  \qquad\qquad  \qquad\quad\quad\;\; \nonumber \\[2mm] 
 & = & \int \frac{d^4p}{(2\pi)^4} \; e^{-\, i\, p\cdot (x_1 - x_2)}\; \frac{[\, C \,(\not\! p + M) \,]_{rs}}{p^2- M^2} \; ,  \\[2mm]
\lefteqn{ (+\, i) \, \big<\!\big< 0 \big| \, \overline{T} [\,  [- \, C\, \psi(x_2)]^+_s  \, [C\, \psi(x_1)]^+_r \,\big| 0 \big> \quad\quad \; = \;  (+\, i) \; \big<\!\big< 0 \big| \, \overline{T} \Big[\, [\,\gamma^0 \, \psi^c(x_2)]_s \,  [\, \overline{\psi(x_1)} \; C\, \gamma^0\, ]_r  \Big] \,\big| 0 \big>  \; =} \nonumber \\[2mm]
 & = & (+\,i) \; \left( \frac{1}{i} \right)^2 \left. \; \frac{\delta^{2}\,  Z_{\tiny \mbox{LWD,0}}[\,\eta\,,\eta^\ast]}{\delta \eta(x_1)_r^\ast \; \delta \eta(x_2)_s^\ast} \right|_{\eta\, =\, \eta^\ast = 0} \; = \; \left( \frac{1}{i} \right)^2 \; \frac{\delta^{2}\,  W_{\tiny \mbox{LWD,0}}[\,\eta\,,\eta^\ast]}{\delta \eta(x_1)_r^\ast \; \delta \eta(x_2)_s^\ast} \nonumber \\[2mm]
 & = & \int \frac{d^4p}{(2\pi)^4} \; e^{+\, i\, p\cdot (x_1 - x_2)}\; \frac{[\, \gamma^0 \, (\not\! p + M^\ast)\, C\,  \gamma^0\, ]_{sr}}{p^2- M^{\ast \, 2}} \nonumber \\[2mm]
 & = & (+\, i) \,\Big( \big<\!\big< 0 \big| \, T [\, \Big[ \,[ C\, \psi(x_1)]_r \, [-\,C\,\psi(x_2)]_s \, \Big] \,\big| 0 \big>\Big)^+ 
\; , \end{eqnarray}
or, equivalently,
\begin{eqnarray} (-\, i) \, \big<\!\big< 0 \big| \, T [\, \psi(x_1) \; \overline{\psi^c(x_2)} \, ] \,\big| 0 \big> & = & \int \frac{d^4p}{(2\pi)^4} \; e^{-\, i\, p\cdot (x_1 - x_2)}\; \frac{\not\! p + M}{p^2- M^2} \; , \\
(+\, i) \; \big<\!\big< 0 \big| \, \overline{T} [\, \psi^c(x_2) \; \overline{\psi(x_1)}\,   ] \,\big| 0 \big> & = & \int \frac{d^4p}{(2\pi)^4} \; e^{+\, i\, p\cdot (x_1 - x_2)}\; \frac{\not\! p + M^\ast}{p^2- M^{\ast \, 2}} \; .
\end{eqnarray} 

\section{The Pais-Uhlenbeck oscillator model and a system of two Klein-Gordon fields}
After having outlined how to appropriately quantize the system of two independent Bosonic Harmonic Oscillators underlying the Pais-Uhlenbeck oscillator model we want illustrate here some non-trivial consequences to Quantum Field Theory arising due to the non-Hermitian nature of the formalism. Starting point will be a comparison of the PT-symmetric action of the two Harmonic Oscillator system as provided in Eq.\ (\ref{lagr3}), i.\ e.,
\begin{eqnarray}\lefteqn{ S_0[Q_1,Q_2,Q^+_1,Q_2^+] \; =} \nonumber \\[2mm]
 & = & \int dt \; \Bigg[ \, \frac{M_1}{2} \left( \dot{Q}_{1} (t)^2 \, - \, \Omega^2_1  \, Q_1(t)^2\right)\quad\;\;\; \; \; + \; \frac{M_2}{2} \left( \dot{Q}_{2} (t)^2 \, - \, \Omega^2_2  \, Q_2(t)^2\right) \nonumber \\
 &  & \quad\;\;\,\, + \; \frac{M^\ast_1}{2} \left( \dot{Q}_{1} (t)^{+\,2} \, - \, \Omega^{\ast\,2}_1  \, Q_1(t)^{+\,2}\right) \; + \; \frac{M^\ast_2}{2} \left( \dot{Q}_{2} (t)^{+\,2} \, - \, \Omega^{\ast\,2}_2  \, Q_2(t)^{+\,2}\right)\; \Bigg] \, ,  
\end{eqnarray}
and the PT-symmetric free action of two Klein-Gordon fields $\varphi_1 (x)$ and $\varphi_2(x)$ described here by 
\begin{eqnarray} S_0[\varphi_1,\varphi_2,\varphi_1^+,\varphi_2^+] \; = \;  \int d^4x & \Bigg[ & \frac{Z_1}{2}\, \Big( (\partial_\mu \varphi_1(x)) (\partial^\mu \varphi_1(x) ) -  M^2_1  \, \varphi_1(x)^2\Big) \nonumber \\
 & + &  \frac{Z_2}{2}\, \Big( (\partial_\mu \varphi_2(x)) (\partial^\mu \varphi_2(x) ) -  M^2_2  \, \varphi_2(x)^2\Big) \nonumber \\[1mm]
 & + & \frac{Z^\ast_1}{2} \, \Big( (\partial_\mu \varphi_1(x)^+) (\partial^\mu \varphi_1(x)^+) -  M^{\ast\,2}_1  \, \varphi_1(x)^{+\,2}\Big) \nonumber \\
 & + & \frac{Z^\ast_2}{2} \, \Big( (\partial_\mu \varphi_2(x)^+) (\partial^\mu \varphi_2(x)^+) -  M^{\ast\,2}_2  \, \varphi_2(x)^{+\,2}\Big) \; \Bigg] \, . 
\end{eqnarray}
It is evident that the role of the mass parameters $M_1$ and $M_2$ in the system of Harmonic oscillators is taken over by the field renormalizations $Z_1$ and $Z_2$ of the Klein-Gordon fields, while the oscillator frequencies $\Omega_1$ and $\Omega_2$ of the Harmonic Oscillators are replaced in the system of Klein-Gordon fields respectively by the differential operators $(-\vec{\nabla}^2 + M^2_1)^{1/2}$ and $(-\vec{\nabla}^2 + M^2_2)^{1/2}$.
This feature gets particularly manifest when comparing the respective Lagrange equations of motion for the Harmonic Oscillator system, i.\ e.,
\begin{eqnarray} & & M_1\left( \ddot{Q}_1 \,  + \,  \Omega^2_1 \;Q_1 \right)   \; = \; 0 \; , \qquad M^\ast_1\left( \ddot{Q}_1^+  \,  + \,  \Omega^{\ast\,2}_1 \; Q_1^+ \right)   \; = \; 0 \; , \nonumber \\[2mm]
 & & M_2\left( \ddot{Q}_2 \,  + \,  \Omega^2_2 \;Q_2 \right)   \; = \; 0 \; , \qquad M^\ast_2\left( \ddot{Q}_2^+  \,  + \,  \Omega^{\ast\,2}_2 \; Q_2^+ \right)   \; = \; 0 \; , 
\end{eqnarray} 
and the system of Klein-Gordon fields, i.\ e.,
\begin{eqnarray} & & Z_1\left( \ddot{\varphi}_1 \,  + \, (-\vec{\nabla}^2 + M^2_1)  \,\varphi_1 \right)   \; = \; 0 \; , \qquad Z^\ast_1\left( \ddot{\varphi}_1^+  \,  + \,   (-\vec{\nabla}^2 + M^{\ast\, 2}_1)  \, \varphi_1^+ \right)   \; = \; 0 \; , \nonumber \\[2mm]
 & & Z_2\left( \ddot{\varphi}_2 \,  + \,   (-\vec{\nabla}^2 + M^2_2)  \,\varphi_2 \right)   \; = \; 0 \; , \qquad Z^\ast_2\left( \ddot{\varphi}_2^+  \,  + \,   (-\vec{\nabla}^2 + M^{\ast\, 2}_2)  \, \varphi_2^+ \right)   \; = \; 0 \; , 
\end{eqnarray} 
In the same way as the energy eigen-values of the quantized system of Harmonic oscillators (see Eq.\ (\ref{spec1})) do not depend on the mass parameters $M_1$ and $M_2$, the energy eigen-values of the quantized system of free Klein-Gordon fields will not depend on the field renormalizations $Z_1$ and $Z_2$ when quantized appropriately \cite{Nakanishi:1972wx}\cite{Nakanishi:1972pt}\cite{Kleefeld:2004jb}\cite{Kleefeld:2004qs}\cite{Kleefeld:2001xd} as specified above. Nonetheless do the  field renormalizations $Z_1$ and $Z_2$ enter the causal and anti-causal retarded Feynman propagators:
\begin{eqnarray}  (-\, i) \, \big<\!\big< 0 \big| \, T [\, \varphi_1(x_1) \, \varphi_1(x_2) \, ] \,\big| 0 \big> \quad & = &  \int \frac{d^4p}{(2\pi)^4} \; e^{-\, i\, p\cdot (x_1 - x_2)}\; \frac{Z^{-1}_1}{p^2- M_1^2} \; ,    \\[2mm] 
(-\, i) \, \big<\!\big< 0 \big| \, T [\, \varphi_2(x_1) \, \varphi_2(x_2) \, ] \,\big| 0 \big> \quad & = &  \int \frac{d^4p}{(2\pi)^4} \; e^{-\, i\, p\cdot (x_1 - x_2)}\; \frac{Z^{-1}_2}{p^2- M_2^2} \; ,    \\[2mm] 
(+\, i) \; \big<\!\big< 0 \big| \, \overline{T} [\, \varphi_1(x_2)^+  \varphi_1(x_1)^+  ] \,\big| 0 \big>  & = &  \int \frac{d^4p}{(2\pi)^4} \; e^{+\, i\, p\cdot (x_1 - x_2)}\; \frac{Z^{\ast\, -1}_1}{p^2- M_1^{\ast \, 2}}  \; , \\   
(+\, i) \; \big<\!\big< 0 \big| \, \overline{T} [\, \varphi_2(x_2)^+  \varphi_2(x_1)^+  ] \,\big| 0 \big>  & = &  \int \frac{d^4p}{(2\pi)^4} \; e^{+\, i\, p\cdot (x_1 - x_2)}\; \frac{Z^{\ast\, -1}_2}{p^2- M_2^{\ast \, 2}}  \; .    
\end{eqnarray}
Despite the fact that the quantization of the system does not depend on the field renormalizations $Z_1$ and $Z_2$ we have pointed out that the respective Minkowskian path integral requires the mass parameters $M_1$ and $M_2$ corresponding in the Harmonic Oscillator system roughly to the oscillator frequencies $\Omega_1$ and $\Omega_2$ to possess some at least infinitesimal negative imaginary part in order to render the  Minkowskian path integral well defined.

The particular feature of the Pais-Uhlenbeck oscillator model under consideration e.\ g.\ by P.~D.~Mannheim \cite{Mannheim:2019aap} is that the mass parameters of the Harmonic Oscillator system $M_1$ and $M_2$ possess opposite sign in analogy to a corresponding Quantum Field Theory describing Quantum Gravity possessing fields where the field renormalizations $Z_1$ and $Z_2$ have opposite sign. In the reference \cite{Mannheim:2019aap} of P.\ D.\ Mannheim there is considered a system, where the field renormalizations are chosen to be $Z_1=+1$ and $Z_2=-1$. 

At the first sight it might appear a rather peculiar situation of the system under consideration by P.\ D.\ Mannheim that the field renormalizations of several fields have opposite signs. Nonetheless it should be pointed out that due to the Lorentz covariance of the action it is a general feature of e.\ g. vector-fields $A^\mu(x)$ that the the field renormalization of the component $A^0(x)$ is entering the action with opposite sign as compared to the further components $\vec{A}(x)$ as the metric induces a sign change e.\ g.\ in the Lorentz scalars like $A_\mu\, A^\mu = A^0A^0 -\vec{A}\cdot \vec{A}$. This can be seen  e.\ g.\ by inspection of the ansatz of St\"uckelberg \cite{Ruegg:2003ps} for the (holomorphic) free action of a massive charged vector field (adapted to our notation as $A^\nu_\pm=(A_1^\mu \pm i A^\mu_2)/\sqrt{2}\,$), i.\ e.:
\begin{eqnarray} S_{0 H} [A_+^\mu,A_-^\mu] & = & \int d^4x \;\left[ -\,  \Big(\partial_\mu\, A_{+\,\nu}(x)\Big) \Big(\partial^\mu A_-^\nu(x)\Big)+M^2\;A_{+\,\nu}(x)A_-^\nu(x)  \right] \nonumber \\[2mm]
& = & \int d^4x \;\;\frac{1}{2}\,\Big[ -\,  \Big(\partial_\mu\, A_{1\,\nu}(x)\Big) \Big(\partial^\mu A_1^\nu(x)\Big)+M^2\;A_{1\,\nu}(x)A_1^\nu(x)   \nonumber \\[2mm]
&  & \qquad\quad\quad\,\;\;  -\,  \Big(\partial_\mu \,A_{2\,\nu}(x)\Big) \Big(\partial^\mu A_2^\nu(x)\Big)+M^2\;A_{2\,\nu}(x)A_2^\nu(x)  \Big] \nonumber \\[2mm]
& = & \int d^4x \;\;\frac{1}{2}\,\Big[ \,-  \Big(\partial_\mu \,\dot{A}^0_1(x)\Big) \Big(\partial^\mu \dot{A}^0_1(x)\Big)\;\;\,+M^2\;A^0_1(x)\,A^0_1(x)   \nonumber \\
&  & \qquad\quad\quad\,\;\;  +\,  \Big(\partial_\mu \,\vec{A}_{1}(x)\Big) \cdot \Big(\partial^\mu \vec{A}_1(x)\Big)-M^2\;\vec{A}_{1}(x)\cdot \vec{A}_2(x)   \nonumber \\[2mm]
&  & \qquad\quad\quad\,\;\;  -\,\Big(\partial_\mu \,\dot{A}^0_2(x)\Big) \Big(\partial^\mu \dot{A}^0_2(x)\Big)\;\;\,+M^2\;A^0_2(x)\,A^0_2(x)   \nonumber \\[2mm]
&  & \qquad\quad\quad\,\;\;  +\,  \Big(\partial_\mu\, \vec{A}_{2}(x)\Big) \cdot \Big(\partial^\mu \vec{A}_2(x)\Big)-M^2\;\vec{A}_{2}(x)\cdot \vec{A}_2(x)  \Big] \; .
\end{eqnarray}
In order to understand the implications of the sign choice for the field renormalizations $Z_1$ and $Z_2$ we consider now shortly the system of the fields $\varphi_1$ and $\varphi_2$ interacting via a coupling constants $g_1$ and $g_2$ with a third Klein-Gordon field $\phi$ with field renormalization $Z=1$ and mass $M$. I.\ e.\ we consider the following action:

\begin{eqnarray} \lefteqn{S[\varphi_1,\varphi_2,\phi,\varphi_1^+,\varphi_2^+,\phi^+ ] \; =} \nonumber \\[2mm]
 & = &  \int d^4x \; \Bigg[ \,\; \; \frac{1}{2}\,\; \Big( (\partial_\mu \phi(x)) (\partial^\mu \phi(x) ) -  M^2  \, \phi(x)^2\Big)  \nonumber \\
 &  & \qquad\;\; \,+\,\frac{Z_1}{2}\, \Big( (\partial_\mu \varphi_1(x)) (\partial^\mu \varphi_1(x) ) -  M^2_1  \, \varphi_1(x)^2\Big)  \, - \, \frac{g_1}{2} \;\phi(x)^2  \; \varphi_1(x) \nonumber \\[1mm]
 &  & \qquad\;\; \,+\,\frac{Z_2}{2}\, \Big( (\partial_\mu \varphi_2(x)) (\partial^\mu \varphi_2(x) ) -  M^2_2  \, \varphi_2(x)^2\Big) \, - \, \frac{g_2}{2} \; \phi(x)^2 \; \varphi_2(x)\nonumber \\[1mm]
 &  & \qquad\;\; \,+\,\; \frac{1}{2} \,\; \Big( (\partial_\mu \phi(x)^+) (\partial^\mu \phi(x)^+) -  M^{\ast \, 2}  \, \phi(x)^{+\, 2}\Big)  \nonumber \\[2mm]
 &  & \qquad\;\; \,+\,  \frac{Z^\ast_1}{2} \, \Big( (\partial_\mu \varphi_1(x)^+) (\partial^\mu \varphi_1(x)^+) -  M^{\ast\,2}_1  \, \varphi_1(x)^{+\,2}\Big)\, - \, \frac{g^\ast_1}{2} \; \varphi_1(x)^+ \, \phi(x)^{+\,2}\nonumber \\[1mm] 
 &  & \qquad\;\; \,+\, \frac{Z^\ast_2}{2} \, \Big( (\partial_\mu \varphi_2(x)^+) (\partial^\mu \varphi_2(x)^+) -  M^{\ast\,2}_2  \, \varphi_2(x)^{+\,2}\Big) \, - \, \frac{g^\ast_2}{2} \; \varphi_2(x)^+\, \phi(x)^{+\,2}\ \; \Bigg] \, . \quad \label{actphiphi1}
\end{eqnarray}
The scattering amplitude $T_{fi}^{\,(2)}$ in second order perturbation theory for the scattering process $\phi(p_1)+\phi(p_2)\rightarrow \phi(p^\prime_1)+\phi(p^\prime_2)$ will be calculated as follows:

\begin{eqnarray}\lefteqn{ i\, T_{fi}^{\,(2)}(p^\prime_1, p^\prime_2; p_1,p_2) \; =} \nonumber \\[2mm]
 & = &  \; \frac{i^2}{2\,!}\;\big<\!\big<0\big| \, a(\vec{p}^{\;\prime}_2)\; a(\vec{p}^{\;\prime}_1) \; T \left[ \left( - \, \frac{g_1}{2} \;\phi(0)^2  \; \varphi_1(0)  \right) \int d^4 x \left(- \, \frac{g_1}{2} \;\phi(x)^2  \; \varphi_1(x) \right)\right]   c^+(\vec{p}_1)\; c^+(\vec{p}_2) \; \big|0\big>_c \nonumber \\
 & + &  \; \frac{i^2}{2\,!}\;\big<\!\big<0\big| \, a(\vec{p}^{\;\prime}_2)\; a(\vec{p}^{\;\prime}_1) \; T \left[ \left( - \, \frac{g_2}{2} \;\phi(0)^2  \; \varphi_2(0)  \right) \int d^4 x \left(- \, \frac{g_2}{2} \;\phi(x)^2  \; \varphi_2(x) \right)\right]   c^+(\vec{p}_1)\; c^+(\vec{p}_2) \; \big|0\big>_c \nonumber \\[2mm]
 & = &  \; \frac{i^2}{2\,!}\int d^4 x\;\left( - \, \frac{g_1}{2} \right)^2\big<\!\big<0\big|  \, T \Big[  \, \varphi_1(0)  \; \varphi_1(x) \, \Big]\, \big|0\big>\big<\!\big<0\big| \, a(\vec{p}^{\;\prime}_2)\; a(\vec{p}^{\;\prime}_1) \; \phi(0)^2 \,\phi(x)^2 \;   c^+(\vec{p}_1)\; c^+(\vec{p}_2) \; \big|0\big>_c 
 \nonumber \\
 & + &  \;   \frac{i^2}{2\,!}\int d^4 x\;\left( - \, \frac{g_2}{2} \right)^2\big<\!\big<0\big|  \, T \Big[  \, \varphi_2(0)  \; \varphi_2(x) \, \Big]\, \big|0\big>\big<\!\big<0\big| \, a(\vec{p}^{\;\prime}_2)\; a(\vec{p}^{\;\prime}_1) \; \phi(0)^2 \,\phi(x)^2 \;   c^+(\vec{p}_1)\; c^+(\vec{p}_2) \; \big|0\big>_c  \nonumber \\[2mm]
 & = & \frac{i^2}{2\,!}\int d^4 x \!\int \frac{d^4p}{(2\pi)^4} \; e^{ i\, p\cdot x} \nonumber \\[2mm]
 & & \cdot\; \left[\, \frac{g^2_1}{4} \;   \frac{i \, Z_1^{-1}}{p^2- M_1^2}+\frac{g^2_2}{4}\;   \frac{i \, Z_2^{-1}}{p^2- M_2^2} \, \right]\; \big<\!\big<0\big| \, a(\vec{p}^{\;\prime}_2)\; a(\vec{p}^{\;\prime}_1) \; \phi(0)^2 \,\phi(x)^2 \;   c^+(\vec{p}_1)\; c^+(\vec{p}_2) \; \big|0\big>_c \; . \label{scattamp1}
\end{eqnarray} 
Assuming without loss of generality (and just for illustration) $g_1^2$ and $g^2_2$ to be real-valued and positive one can see that the choice  $Z_1=+1$ and $Z_2=-1$ will lead to the fact that the scattering amplitude containing the exchange of the $\varphi_2$-field will enter --- in the spirit of the formalism of W.\ Pauli and F.\ Villars \cite{Pauli:1949zm}\cite{Schweber:1994qa} --- with an opposite overall sign as compared to the scattering amplitude containing the exchange  of the $\varphi_1$-field.

One could now be tempted to remove the effect of the field renormalizations $Z_1$ and $Z_2$ by replacing the fields $\varphi_1$ and $\varphi_2$ via introducing new Klein-Gorden fields $\phi_1$ and $\phi_2$ defined in the following way:
\begin{equation} \phi_1 = Z^{\,\frac{\,1}{2}}_1 \, \varphi_1 \; , \qquad \phi_2 = Z^{\,\frac{1}{2}}_2 \, \varphi_2 \qquad \Leftrightarrow \qquad \varphi_1 = Z^{-\frac{\,1}{2}}_1 \, \phi_1 \; , \qquad \varphi_2 = Z^{-\frac{1}{2}}_2 \, \phi_2 \; . \end{equation}
In applying this replacement to the action $S[\varphi_1,\varphi_2,\phi,\varphi_1^+,\varphi_2^+,\phi^+ ]$ in Eq.\ (\ref{actphiphi1}) we obtain the following new action $S[\phi_1,\phi_2,\phi,\phi_1^+,\phi_2^+,\phi^+ ]$:

\begin{eqnarray} \lefteqn{S[\phi_1,\phi_2,\phi,\phi_1^+,\phi_2^+,\phi^+ ] \; =} \nonumber \\[2mm]
 & = &  \int d^4x \; \Bigg[ \,\;  \frac{1}{2}\,\Big( (\partial_\mu \phi(x)) (\partial^\mu \phi(x) ) -  M^2  \, \phi(x)^2\Big)  \nonumber \\
 &  & \qquad\;\; \,+\,\frac{1}{2}\, \Big( (\partial_\mu \phi_1(x)) (\partial^\mu \phi_1(x) ) -  M^2_1  \, \phi_1(x)^2\Big)  \, - \, \frac{g_1}{2\,Z^{1/2}_1} \;  \phi(x)^2  \; \phi_1(x) \nonumber \\[1mm]
 &  & \qquad\;\; \,+\,\frac{1}{2}\, \Big( (\partial_\mu \phi_2(x)) (\partial^\mu \phi_2(x) ) -  M^2_2  \, \phi_2(x)^2\Big) \, - \, \frac{g_2}{2\,Z^{1/2}_2}  \; \phi(x)^2 \; \phi_2(x)\nonumber \\
 &  & \qquad\; \,+\, \frac{1}{2} \,\; \Big( (\partial_\mu \phi(x)^+) (\partial^\mu \phi(x)^+) -  M^{\ast \, 2}  \, \phi(x)^{+\, 2}\Big)  \nonumber \\[2mm]
 &  & \qquad\;\; \,+\,  \frac{1}{2} \, \Big( (\partial_\mu \phi_1(x)^+) (\partial^\mu \phi_1(x)^+) -  M^{\ast\,2}_1  \, \phi_1(x)^{+\,2}\Big)\, - \, \frac{g^\ast_1}{2\,Z^{\ast\, 1/2}_1} \; \phi_1(x)^+ \, \phi(x)^{+\,2}\nonumber \\
 &  & \qquad\;\;\,+\, \frac{1}{2} \, \Big( (\partial_\mu \phi_2(x)^+) (\partial^\mu \phi_2(x)^+) -  M^{\ast\,2}_2  \, \phi_2(x)^{+\,2}\Big) \, - \, \frac{g^\ast_2}{2\,Z^{\ast\, 1/2}_2} \; \phi_2(x)^+\, \phi(x)^{+\,2}\ \; \Bigg] \, . \quad \quad\label{actphiphi2}
\end{eqnarray}
At first sight the new action appears more convenient than the original action due to trivial field  renormalization constants of all Klein-Gordon fields. At the second sight one has to notice that the effect of the original field renormalization constants $Z_1$ and $Z_2$ has shifted to a rescaling of the coupling constants $g_1$ and $g_2$  by the factors $Z^{-1/2}_1$ and $Z^{-1/2}_2$ respectively which will be purely imaginary, if one of the constants  $Z_1$ and $Z_2$ turns out to be real-valued and negative. Hence, the non-Hermitian nature of the system gets now manifest in the coupling constants which turn out to be complex-valued. Clearly, as pointed out before, the whole system is already non-Hermitian due to the requirement that the mass parameters $M$, $M_1$ and $M_2$  have to possess an at least infinitesimal imaginary part to render the Minkowskian path integral well defined.

For comparison the scattering amplitude for the processes $\phi(p_1)+\phi(p_2)\rightarrow \phi(p^\prime_1)+\phi(p^\prime_2)$  can be calculated --- with an identical result --- analogously to Eq.\ (\ref{scattamp1}) in terms of the new fields $\phi$, $\phi_1$ and $\phi_2$:

\begin{eqnarray}\lefteqn{ i\, T_{fi}^{\,(2)}(p^\prime_1, p^\prime_2; p_1,p_2) \; =} \nonumber \\[2mm]
 & = &  \; \frac{i^2}{2\,!}\;\int d^4 x \nonumber \\[2mm]
 & & \bigg\{\big<\!\big<0\big| \, a(\vec{p}^{\;\prime}_2)\; a(\vec{p}^{\;\prime}_1) \; T \left[ \left( - \, \frac{g_1}{2\,Z^{1/2}_1} \phi(0)^2   \phi_1(0)  \right)  \left(- \, \frac{g_1}{2\,Z^{1/2}_1} \;\phi(x)^2  \; \phi_1(x) \right)\right]   c^+(\vec{p}_1)\; c^+(\vec{p}_2) \, \big|0\big>_c \nonumber \\
 &  &  + \big<\!\big<0\big| \, a(\vec{p}^{\;\prime}_2)\; a(\vec{p}^{\;\prime}_1) \; T \left[ \left( - \, \frac{g_2}{2\,Z^{1/2}_2} \phi(0)^2  \phi_2(0)  \right) \left(- \, \frac{g_2}{2\,Z^{1/2}_2} \;\phi(x)^2  \; \phi_2(x) \right)\right]   c^+(\vec{p}_1)\; c^+(\vec{p}_2) \, \big|0\big>_c     \bigg\}\nonumber \\[2mm]
 & = &  \; \frac{i^2}{2\,!}\;\int d^4 x \nonumber \\[2mm]
 & & \bigg\{ \left( - \, \frac{g_1}{2\,Z^{1/2}_1} \right)^2\big<\!\big<0\big|  \, T \Big[  \, \phi_1(0)  \; \phi_1(x) \, \Big]\, \big|0\big>\big<\!\big<0\big| \, a(\vec{p}^{\;\prime}_2)\; a(\vec{p}^{\;\prime}_1) \; \phi(0)^2 \,\phi(x)^2 \;   c^+(\vec{p}_1)\; c^+(\vec{p}_2) \; \big|0\big>_c 
 \nonumber \\
 &  &  +  \left( - \, \frac{g_2}{2\,Z^{1/2}_2} \right)^2 \big<\!\big<0\big|  \, T \Big[  \, \phi_2(0)  \; \phi_2(x) \, \Big]\, \big|0\big>\big<\!\big<0\big| \, a(\vec{p}^{\;\prime}_2)\; a(\vec{p}^{\;\prime}_1) \; \phi(0)^2 \,\phi(x)^2 \;   c^+(\vec{p}_1)\; c^+(\vec{p}_2) \; \big|0\big>_c  \bigg\}\nonumber \\[2mm]
 & = & \frac{i^2}{2\,!}\int d^4 x \!\int \frac{d^4p}{(2\pi)^4} \; e^{ i\, p\cdot x} \nonumber \\[2mm]
 & & \cdot\; \left[\, \frac{g^2_1}{4\,Z_1} \;   \frac{i}{p^2- M_1^2}+\frac{g^2_2}{4\,Z_2}\;   \frac{i}{p^2- M_2^2} \, \right]\; \big<\!\big<0\big| \, a(\vec{p}^{\;\prime}_2)\; a(\vec{p}^{\;\prime}_1) \; \phi(0)^2 \,\phi(x)^2 \;   c^+(\vec{p}_1)\; c^+(\vec{p}_2) \; \big|0\big>_c \; . \label{scattamp2}
\end{eqnarray} 
The comparison of this result with the corresponding result displayed in Eq.\ (\ref{scattamp1})  shows that the effect of the renormalization constants $Z_1$ and $Z_2$ has shifted from the propagators of the fields $\varphi_1$ and $\varphi_2$ to the interactions of the respective fields $\phi_1$ and $\phi_2$. 
\section{Conclusions}
Throughout this manuscript we have tried to convince the open-minded reader that Quantum Theory, i.\ e., Quantum Mechanics and Quantum Field Theory, should be considered to be inherently non-Hermitian to avoid serious inconsistencies arising in a naive Hermitian framework. The general tool to avoid these inconsistencies as prescribed already at various places (see e.\ g.\  \cite{Kleefeld:2004jb}\cite{Kleefeld:2002au}) we are going to call from now on {\em PT-symmetric completion}. It implies that before quantization there has to be added to any Lagrange function $L_H$ (being called here holomorphic Lagrange function) --- containing degrees of freedom which should be considered to be complex-valued --- the Hermitian conjugate Lagrange function $L_A=L_H^+$ (called here anti-holomorphic Lagrange function) in order to obtain an overall PT-symmetric Lagrange function $L=L_H+L_A$ containing now twice as many independent degrees of freedom as the original holomorphic Lagrange function $L_H$. After quantization of the PT-symmetric system the PT-symmetric Hamilton operator $H=H_H+H_A$ will not only contain the spectrum of the holomorphic Hamilton operator $H_H$, yet also the complex conjugate spectrum of the anti-holomorphic Hamilton operator $H_A=H^+_H$.

The original motivation for the PT-symmetric completion as described e.\ g.\ in \cite{Kleefeld:2004jb}\cite{Kleefeld:2002au} has been to avoid violations of causality, analyticity and Lorentz invariance of the formalism. In the same way, as the holomorphic Hamilton operator $H_H$ has been associated to the evolution of a system under consideration with causal boundary conditions described by holomorphic functions, the anti-holomorphic Hamilton operator $H_A$ has been associated to the evolution of the system with anti-causal boundary conditions described by anti-holomorphic functions. The afore-mentioned PT-symmetric completion avoids any interaction between causal and anti-causal degrees of freedom which would lead to causality and analyticity violations.

The price to be paid by performing a PT-symmetric completion of a physical system is that Feynman path integrals, generating functionals and vacuum persistence amplitudes have to be generalized in the way as sketched in the present manuscript for quantum mechanical systems like the simple Harmonic Oscillator and for quantum field theoretic systems with Nakanishi-Klein-Gordon and Lee-Wick-Dirac fields.

As demonstrated throughout the manuscript the conjecture about an eventual divergence of the norm  $\big<\!\big<0\big|0\big>$ of the vacuum state as pointed out by P.\ D.\ Mannheim \cite{Mannheim:2023ivd} is not an issue when quantizing a physical system appropriately within a non-Hermitian framework along suitable complex-valued integration contours.

Furthermore we have tried to argue by using the example of Lorentz scalars of vector fields like e.~g.\ $A_\mu A^\mu = A^0 \, A^0 - \vec{A}\cdot \vec{A} = - A_E^0\, A_E^0 - \vec{A}\cdot \vec{A}$ that the change of a Minkoskian to a Euclidean description of the system on the level of fields will shift the underlying non-Hermiticity of the formalism self-consistently to the interactions of Euclidean fields like e.\ g.\ $A^0_E\, $. It is important to note that this shift {\em cannot} be performed consistently in a naive Hermitian formalism of Quantum Field Theory.

Interestingly we had observed the appearance of non-Hermitian interactions already in 2002~\cite{Kleefeld:2002au} when replacing the seemingly Hermitian Lagrangian of Quantum-Chromo-Dynamics (QCD) by a manifestly non-Hermitian  Quark-Level-Linear-Sigma-Model Lagrangian allowing to construct --- contrary to what has been stated for a long time \cite{Kleefeld:2005hf} --- an asymptotically free theory of strong interactions not only on the basis of vector confinement, yet also on the basis of scalar confinement.

In Eqs.\ (\ref{scattamp1}) and (\ref{scattamp2}) we have considered the scattering process $\phi(p_1)+\phi(p_2)\rightarrow \phi(p^\prime_1)+\phi(p^\prime_2)$ proceeding via the virtual exchange of two Klein-Gordon fields $\varphi_1$ and $\varphi_2$ (or $\phi_1$ and $\phi_1$) with mass $M_1$ and $M_2$ respectively.  The scattering amplitude in second order perturbation theory turned out to be proportional to the following term:
\begin{equation}     \frac{g^2_1\,  Z_1^{-1} }{p^2- M_1^2}\; +\;   \frac{g^2_2\, Z_2^{-1}}{p^2- M_2^2} \; = \; \frac{p^2 \, \Big(g^2_1 \,Z_1^{-1} +g^2_2 \,Z_2^{-1}\Big)\, - \, \Big(M^2_2 \, g^2_1 \,Z_1^{-1} + M^2_1\, g^2_2 \,Z_2^{-1}\Big)}{(p^2- M_1^2)(p^2- M_2^2)}\; . \label{combprop1}
\end{equation}
Due to the PT-symmetric completion of the system the mass parameters $M_1$ and $M_2$, the renormalization parameters $Z_1$ and $Z_2$ and the coupling constants $g_1$ and $g_2$  can be chosen to be complex-valued. Keeping this feature in mind we have many ways to realize the constraint 
\begin{equation} g^2_1 \,Z_1^{-1} +g^2_2 \,Z_2^{-1} \; = \; 0 \; ,
\end{equation}
allowing to increase the power behaviour in the denominator in the combined propagator as suggested e.\ g.\ in the formalism of W.\ Pauli and F.\ Villars \cite{Pauli:1949zm} to improve the convergence of Quantum Field Theory or by P.\ D.\ Mannheim \cite{Mannheim:2023ivd} in order to provide the renormalizability of Quantum Gravity. In our non-Hermitian formalism the fulfillment of the constraint cannot only be achieved by choosing --- for real-valued coupling constants $g_1$ and $g_2$ --- renormalization parameters $Z_1$ and $Z_2$ of opposite sign, yet also by choosing e.\  g.\ --- for real-valued positive renormalization parameters $Z_1$ and $Z_2$ --- the squares of coupling constants $g^2_1$ and $g^2_2$ to be of opposite sign implying one of the couplings e.\ g. to be real-valued and the other coupling to be purely imaginary. As a side remark we would like raise attention to the fact   that the gluon propagator determined in QCD on the basis of Schwinger-Dyson equations is also positivity violating and displaying an abnormal power  behaviour as a function of the four-momentum square \cite{Strauss:2012zz}.

A particularly interesting situation will occur when the following two constraints will be fulfilled simultaneously implying a vanishing scattering amplitude:
\begin{equation} g^2_1 \,Z_1^{-1} +g^2_2 \,Z_2^{-1} \; = \; 0 \quad \mbox{and} \quad M^2_2 \, g^2_1 \,Z_1^{-1} + M^2_1\, g^2_2 \,Z_2^{-1} \; = 0\; .
\end{equation}
At first sight this situation appears rather academic or mind-boggling, since it shows by inspection of Eq.\ (\ref{combprop1}) that a physical system can contain two ``spurious" fields whose contributions to scattering processes seem to cancel exactly. Nonetheless could it be very fruitfully used to explain the also mind-boggling contradiction that the Lorentz covariant vector field $A^\mu$ of the photon possessing four-components $A^0$, $A^1$, $A^2$  and $A^3$ appears in Nature only with two transverse components $A^1$ and $A^2$ while the components $A^0$ and $A^3$ seem to be --- in the spirit of S.\ N.\ Gupta \cite{Gupta:1949rh} and K.~Bleuler \cite{Bleuler:1950cy} --- invisible or ``spurious" in the above sense.

The discussion in this manuscript has tried to illustrate that the PT-completion of  Quantum Theory being accessible to standard and new renormalization techniques  is opening many new doors for open-minded scientists in order to surpass formalistic obstacles in the description of Nature posed by the naive and unreflected use of Hermitian Quantum Theory. 

 In order to test the formalism of PT-completion we encourage the interested reader to perform the quantization of the fourth-order derivative Pais-Uhlenbeck  oscillator model Eqs.\ (\ref{lagpaisuhl1}) or (\ref{lagbenman1}) after PT-completion with the help of the formalism of M.~Ostrogradski \cite{deUrries:1998obu}\cite{Smilga:2017arl} for higher derivative scalar field theories. The resulting spectrum of this exercise  will turn out to be identical to the spectrum of the PT-symmetric system of two independent Bosonic Harmonic Oscillators provided in Eq.~(\ref{spec1}), which has been obtained here by standard canonical quantization.

\ack This work is dedicated to the pupils of my present school and working place ``Wolfgang-Borchert Gymnasium" in Langenzenn, Germany, who do not stop reminding their exhausted physics and mathematics teacher every day to publish his accumulated research ideas due to their certain belief that they will have some relevance to the understanding of Nature.


\section*{References}
\begin{thebibliography}{110}

\bibitem{Mannheim:2019aap}
Mannheim P D 2019 Is dark matter fact or fantasy? \textemdash{} Clues from the data \\
{\it Int. J. Mod. Phys.} D {\bf 28}  1944022 ({\it Preprint} [arXiv:1903.11217 [astro-ph.GA]])

\bibitem{Mannheim:2023ivd}
Mannheim P D 2023 Determining the normalization of the quantum field theory vacuum, with implications for quantum gravity
{\it Preprint} [arXiv:2301.13029 [hep-th]]

\bibitem{deUrries:1998obu}
De Urries F J and Julve J 1998
Ostrogradski formalism for higher derivative scalar field theories\\
{\it J. Phys.} A {\bf 31} 6949
({\it Preprint} [arXiv:hep-th/9802115 [hep-th]])

\bibitem{Smilga:2017arl}
Smilga A 2017
Classical and quantum dynamics of higher-derivative systems\\
{\it Int. J. Mod. Phys.} A {\bf 32} 1730025
({\it Preprint} [arXiv:1710.11538 [hep-th]])

\bibitem{Mandal:2022xuw}
Mandal B P, Pandey V K and Thibes R 2022 BFV quantization and BRST symmetries of the gauge invariant fourth-order Pais-Uhlenbeck oscillator
{\it Nucl. Phys.} B {\bf 982} 115905
({\it Preprint} [arXiv:2202.00715 [hep-th]])

\bibitem{Bender:2007wu}
Bender C M and Mannheim P D 2008
No-ghost theorem for the fourth-order derivative Pais-Uhlenbeck oscillator model 
{\it Phys. Rev. Lett.} {\bf 100} 110402
({\it Preprint} [arXiv:0706.0207 [hep-th]])

\bibitem{Pais:1950za}
Pais A and Uhlenbeck G E 1950 On Field theories with nonlocalized action
{\it Phys. Rev.} {\bf 79} 145

\bibitem{Mannheim:2004qz}
Mannheim P D and Davidson A 2005
Dirac quantization of the Pais-Uhlenbeck fourth order oscillator \\
{\it Phys. Rev.} A {\bf 71}, 042110
({\it Preprint} [arXiv:hep-th/0408104 [hep-th]])

\bibitem{Fluegge:1994}
Fl\"ugge S 1994 {\it Practial quantum mechanics} (2nd printing) (Berlin, Heidelberg, New York: Springer)\\
ISBN 3-450-07050-8

\bibitem{Kleefeld:2003zj}
Kleefeld F 2003 On symmetries in (anti)causal (non)Abelian quantum theories
{\it eConf} {\bf C0306234} 1367 and 2004 {\it Proc. Inst. Math. NAS Ukraine} {\bf 50} 1367
({\it Preprint} [arXiv:hep-th/0310204 [hep-th]])

\bibitem{Bender:2007nj}
Bender C M 2007 Making sense of non-Hermitian Hamiltonians\\
{\it Rept. Prog. Phys.} {\bf 70} 947
({\it Preprint} [arXiv:hep-th/0703096 [hep-th]])

\bibitem{Bender:2004sv}
Bender C M, Brandt S F, Chen J H and Wang Q h 2005 Ghost busting: PT-symmetric interpretation of the Lee model
{\it Phys. Rev.} D {\bf 71} 025014 
({\it Preprint} [arXiv:hep-th/0411064 [hep-th]])

\bibitem{Bender:1998ke}
Bender C M and Boettcher S 1998 Real spectra in non-Hermitian Hamiltonians having PT symmetry\\
{\it Phys. Rev. Lett.} {\bf 80} 5243
({\it Preprint} [arXiv:physics/9712001 [physics]])

\bibitem{Weigert:2003py}
Weigert S 2003 Completeness and orthonormality in PT-symmetric quantum systems\\
{\it Phys. Rev.} A {\bf 68} 062111
({\it Preprint} [arXiv:quant-ph/0306040 [quant-ph]])

\bibitem{Znojil:2001xy}
Znojil M 2004 Conservation of pseudo-norm in PT symmetric quantum mechanics\\ 
{\it Rend. Circ. Mat. Palermo, Serie II, Suppl.} {\bf 72} 211
({\it Preprint} [arXiv:math-ph/0104012 [math-ph]])

\bibitem{Nakanishi:1972wx}
Nakanishi N 1972 Covariant formulation of the complex-ghost relativistic field theory and the Lorentz noninvariance of the S matrix
{\it Phys. Rev.} D {\bf 5} 1968

\bibitem{Nakanishi:1972pt}
Nakanishi N 1972 Indefinite metric quantum field theory
{\it Prog. Theor. Phys. Suppl.} {\bf 51} 1

\bibitem{Froissart:1959occ}
Froissart M 1959 Covariant formalism of a field with indefinite metric
{\it Nuovo Cim. Suppl.} {\bf 14} 197

\bibitem{Kleefeld:2004jb}
Kleefeld F 2004 Non-Hermitian quantum theory and its holomorphic representation: Introduction and some applications
{\it Preprint} [arXiv:hep-th/0408028 [hep-th]]

\bibitem{Kleefeld:2004qs}
Kleefeld F 2004 Non-Hermitian quantum theory and its holomorphic representation: Introduction and applications
{\it Preprint} [arXiv:hep-th/0408097 [hep-th]]  (Contribution to the 2nd Int.\ Workshop on Pseudo-Hermitian Hamiltonians in Quantum Physics (2004, June 14-16, Villa Lanna, Prague, Czech Republic))

\bibitem{Kleefeld:2002au}
Kleefeld F 2003 Consistent relativistic quantum theory for systems / particles described by non-Hermitian Hamiltonians and Lagrangians
{\it AIP Conf. Proc.} {\bf 660} no.1, 325
({\it Preprint} [arXiv:hep-ph/0211460 [hep-ph]])

\bibitem{Kleefeld:2002gw}
Kleefeld F 2003 Does it make any sense to talk about a delta isobar?\\
{\it Few Body Syst. Suppl.} {\bf15} 201
({\it Preprint} [arXiv:nucl-th/0212008 [nucl-th]])

\bibitem{Feynman:1948ur}
Feynman R P 1948 Space-time approach to nonrelativistic quantum mechanics
{\it Rev. Mod. Phys.} {\bf 20} 367
\bibitem{Feynman:1949zx}
Feynman R P 1949 Space-time approach to quantum electrodynamics
{\it Phys. Rev.} {\bf 76} 769

\bibitem{Dittrich:1992et}
Dittrich W and Reuter M 1992 {\it Classical and quantum dynamics: from classical paths to path integrals} (Berlin, Heidelberg: Springer)

\bibitem{Greiner:1996zu}
Greiner W and Reinhardt J  1996
{\it Field quantization} (Berlin: Springer)

\bibitem{Ryder:1985wq}
Ryder L H 1985 {\it Quantum field theory} (Cambridge University Press)

\bibitem{Kleinert:1993}
Kleinert H 1993 {\it Pfadintegrale in Quantenmechanik, Statistik und Polymerphysik} (Mannheim, Germany: Bibliographisches Institut \& F.\ A.\ Brockhaus AG)

\bibitem{Bagarello:2020jhq}
Bagarello F and Feinberg J 2020 Bicoherent-state path integral quantization of a non-Hermitian Hamiltonian
{\it Annals Phys.} {\bf 422} 168313
({\it Preprint} [arXiv:2001.04955 [math-ph]])

\bibitem{Economou:1979}
Economou E N 1979 {\it Green's functions in quantum physics} (Springer Series in Solid-State Sciences 7) ed P~Fulde (Berlin, Heidelberg, New York: Springer)

\bibitem{Brush:1961}
Brush S G 1961 Functional integrals and statistical physics
{\it Rev. Mod. Phys.} {\bf 33} 79

\bibitem{Kleefeld:2001xd}
Kleefeld F, van Beveren E and Rupp G 2001 The Pionic width of the $\omega(782)$ meson within a well-defined, unitary quantum field theory of (anti-)particles and (anti-)holes
{\it Nucl. Phys.} A {\bf 694} 470
({\it Preprint} [arXiv:hep-ph/0101247 [hep-ph]])

\bibitem{Kleefeld:1999}
Kleefeld F 1999 {\it Exklusive Schwellenproduktion von $\pi^0$-, $\eta$- und $K^+$-Mesonen in Proton-Proton-St\"o\ss en} (Doctoral thesis, Faculty of (Natural) Sciences, FA-University Erlangen-N\"urnberg, Germany)

\bibitem{Kleefeld:1998dg}
Kleefeld F 1998 Consistent effective description of nucleonic resonances in an unitary relativistic field theoretic way  
{\it Preprint} [arXiv:nucl-th/9811032 [nucl-th]] and 2000 {\it Proc. of the XIV Int. Seminar on High Energy Physics Problems (ISHEPP 1998, August 17-22, Dubna): Relativistic Nuclear Physics and Quantum Chromodynamics} ed A M Baldin and V V Burov (JINR, Dubna, Russian Federation) vol. 1, pp.\ 69-77   

\bibitem{Kleefeld:1998yj}
Kleefeld F 1999 Consistent effective field theoretic treatment of resonances with nonzero width
{\it Acta Phys. Polon.} B {\bf 30} 981
({\it Preprint} [arXiv:nucl-th/9806060 [nucl-th]])

\bibitem{Kleefeld:2021sqz}
Kleefeld F 2021 Identification of the metric for diagonalizable (anti-)pseudo-Hermitian Hamilton operators represented by two-dimensional matrices
{\it Preprint} [arXiv:2102.08182 [quant-ph]]

\bibitem{Lee:1970iw}
Lee T D and Wick G C 1970 Finite theory of quantum electrodynamics
{\it Phys. Rev.} D {\bf 2} 1033

\bibitem{Cotaescu:1983nc}
Cot\v{a}escu I I 1983 Probabilistic interpretation of the dipole ghost models
{\it Phys. Rev.} D {\bf 27} 2556

\bibitem{Ruegg:2003ps}
Ruegg H and Ruiz-Altaba M 2004 The St\"uckelberg field\\
{\it Int. J. Mod. Phys.} A {\bf 19} 3265
({\it Preprint} [arXiv:hep-th/0304245 [hep-th]])

\bibitem{Pauli:1949zm}
Pauli W and Villars F 1949 On the Invariant regularization in relativistic quantum theory\\
{\it Rev. Mod. Phys.} {\bf 21} 434

\bibitem{Schweber:1994qa}
Schweber S S 1994 {\it QED and the men who made it: Dyson, Feynman, Schwinger, and Tomonaga}\\ (Princeton University Press: Princeton, New Jersey) 
ISBN 0-691-03327-7

\bibitem{Kleefeld:2005hf}
Kleefeld F 2006 Kurt Symanzik: A Stable fixed point beyond triviality\\
{\it J. Phys.} A {\bf 39}, L9 
({\it Preprint} [arXiv:hep-th/0506142 [hep-th]])

\bibitem{Strauss:2012zz}
Strauss S, Fischer C S and Kellermann C 2012 Analytic structure of Landau gauge ghost and gluon propagators
{\it Prog. Part. Nucl. Phys.} {\bf 67} 239
({\it Preprint} [arXiv:1208.6239 [hep-ph]])

\bibitem{Gupta:1949rh}
Gupta S N 1950 Theory of longitudinal photons in quantum electrodynamics
{\it Proc. Phys. Soc.} A {\bf 63} 681

\bibitem{Bleuler:1950cy}
Bleuler K 1950 Eine neue Methode zur Behandlung der longitudinalen und skalaren Photonen\\
{\it Helv. Phys. Acta} {\bf 23} 567

\end{thebibliography}
\end{document}